\documentclass[prb,aps,showpacs,twocolumn,floatfix,citeautoscript,superscriptaddress]{revtex4-1}
\usepackage{amsmath,amssymb,physics,color,array}
\usepackage{graphicx}
\usepackage[autostyle]{csquotes}

\newcommand{\Mab}[4]{\left(\begin{array}{cc}
#1\ & #2 \\
#3\ & #4 \end{array} \right)}

\begin{document}

\title{Tunable pseudogaps due to non-local coherent transport\\ in voltage-biased three-terminal Josephson junctions}

\author{C. Padurariu}
\affiliation{Centre National de la Recherche Scientifique, Institut NEEL, F-38042 Grenoble Cedex 9, France}
\affiliation{Universit\'e Grenoble-Alpes, Institut NEEL, F-38042 Grenoble Cedex 9, France}
\affiliation{Low Temperature Laboratory, Department of Applied Physics, Aalto University, P.O. Box 15100, FI-00076 Aalto, Finland}
\author{T. Jonckheere}
\affiliation{Aix Marseille Univ, Universit\'e de Toulon, CNRS, CPT, Marseille, France}
\author{J. Rech}
\affiliation{Aix Marseille Univ, Universit\'e de Toulon, CNRS, CPT, Marseille, France}
\author{T. Martin}
\affiliation{Aix Marseille Univ, Universit\'e de Toulon, CNRS, CPT, Marseille, France}
\author{D. Feinberg}
\affiliation{Centre National de la Recherche Scientifique, Institut NEEL, F-38042 Grenoble Cedex 9, France}
\affiliation{Universit\'e Grenoble-Alpes, Institut NEEL, F-38042 Grenoble Cedex 9, France}

\begin{abstract}

We investigate the proximity effect in junctions between $N=3$ superconductors under 
commensurate voltage bias. The bias is chosen to highlight the role of transport 
processes that exchange multiple Cooper pairs coherently between more than two superconductors. 
Such non-local processes can be studied in the dc response, where local transport processes 
do not contribute.
We focus on the proximity-induced normal density of states that we investigate in a 
wide parameter space. We reveal the presence of deep and highly tunable pseudogaps 
and other rich structures. 
These are due to a static proximity effect that is absent for $N=2$ and is sensitive to an emergent
superconducting phase associated to non-local coherent transport. In comparison with results
for $N=2$, we find similarities in the signature peaks of multiple Andreev reflections. 
We discuss the effect of electron-hole decoherence and of various types of junction asymmetries.
Our predictions can be investigated experimentally using tunneling spectroscopy.

\end{abstract}


\maketitle

\section {Introduction}
\label{introduction}

Quantum transport in Josephson junctions has been the focus of extensive research, 
predominantly studying junctions between $N=2$ superconductors. Short junctions 
exhibit a strong proximity effect that manifests in equilibrium as an induced minigap 
in the density of states. A finite minigap is accompanied by a non-dissipative 
superconducting current. On the contrary, out of equilibrium dynamics due to voltage bias
leads to entirely dissipative quasiparticle transport in two-terminal junctions. 
When the bias voltage $V$ is below the superconducting 
gap of the leads $\Delta$, the dissipative quasiparticle motion is described by 
multiple Andreev reflections (MAR) \cite{MAR}. 
Electrons and holes cross the structure, being 
Andreev-reflected at each junction interface. Each crossing provides the energy 
$eV$, giving rise to features in the $I(V)$ 
curve\cite{Scheer} at integer fractions of $2\Delta/e$.
In the regime dominated by MAR, the density of states no longer manifests a clear minigap, \cite{Pierre, Bardas}
instead exhibiting peaks located at energy intervals separated by $eV$.

\begin{figure}[t]
\centerline{\includegraphics[width=0.9\linewidth]{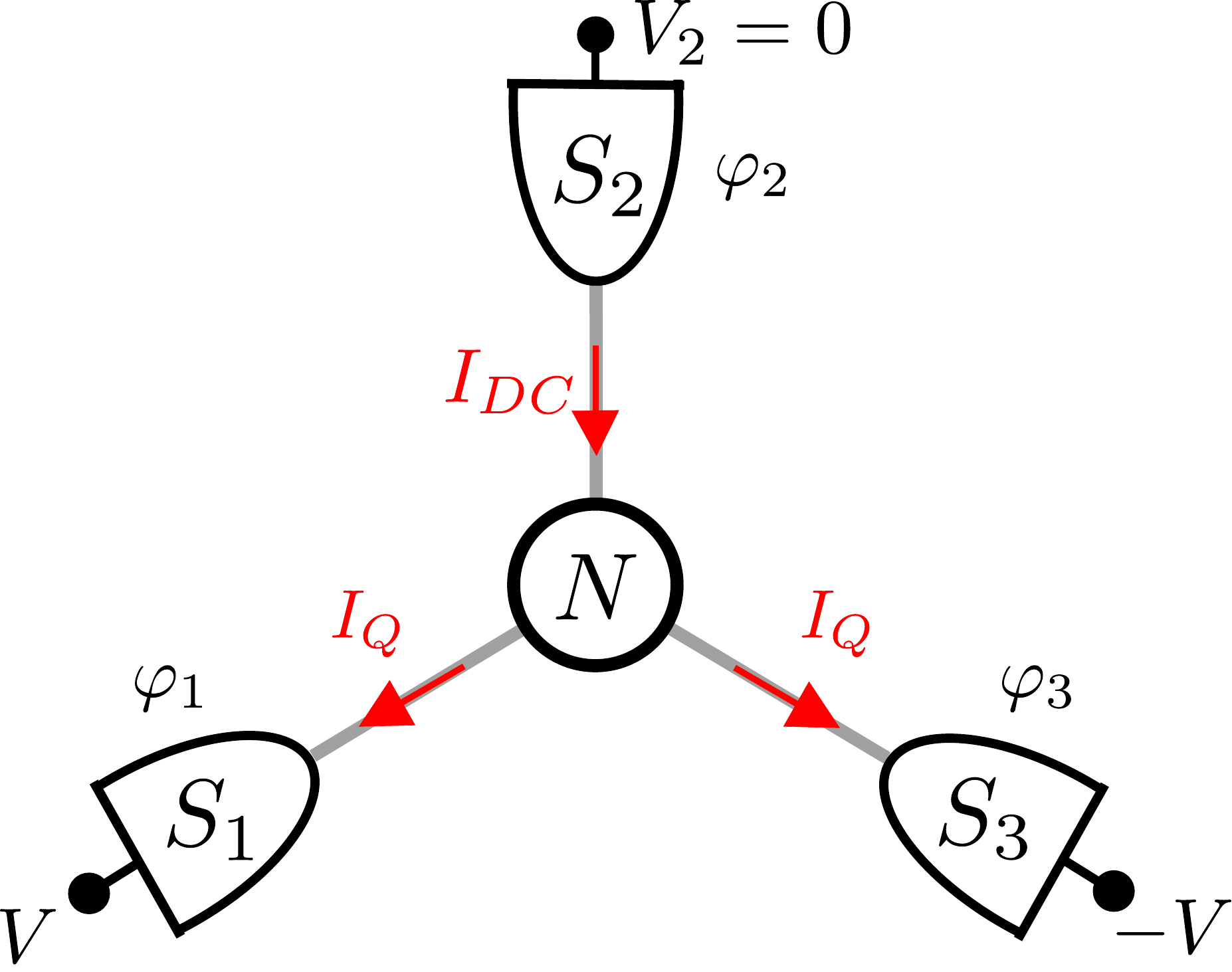}}
\caption{(Color online.) Schematic of the setup. Three superconducting electrodes $S_1$, $S_2$, and $S_3$
at voltages $V_1=V,V_2=0$, and $V_3=-V$ and with superconducting phases $\varphi_1$, $\varphi_2$, and
$\varphi_3$ are connected to a short, diffusive metallic region. The dc current, 
$I_{DC}$, in $S_2$ drives equal dc currents $I_Q$ in $S_1$ and $S_3$, 
as explained in Sec. \ref{phenomenology}.}
\label{fig:setup}
\end{figure}

Recently, the study of junctions between $N\geq 3$ superconductors 
has attracted considerable interest, both theoretical \cite{Cuevas,Houzet,Freyn,Jonckheere} 
and experimental \cite{Pfeffer,Giazotto,Ronen,JeanEude}. 
Unique features have been revealed, that do not manifest in the $N=2$ 
junctions. 
In equilibrium, mapping the subgap 
Andreev spectrum has revealed level crossings at zero energy for non-trivial phase values 
\cite{Akhmerov,Padurariu,Riwar,Giazotto}. For $N\geq 4$, the crossing point was shown to 
have analogous topological properties to Weyl points in topological semi-metals \cite{Riwar,Yulinew}. 

Voltage bias further emphasizes the complex phenomenology of $N\geq 3$ junctions.
When the voltages are chosen such that the ratio of any 
two is a rational number (commensurate bias), 
the transport is no longer entirely dissipative as is the case in $N=2$ junctions.
Previous works have shown that a 
non-dissipative dc current component\cite{Cuevas,Freyn,Jonckheere}
arises in the junction due to coherent exchange of
multiple Cooper pairs non-locally between three or more superconductors. 
The non-local current is sensitive to bias, as well as an emerging \textit{stationary} phase that 
is obtained by combining the phases of multiple superconductors.

The simplest setup consists of the three-terminal Josephson junction (TTJ) where the 
non-dissipative current is expected to be largest when the two independent phases are
affected by opposite voltage bias, $V$ and $-V$, as shown in Fig. \ref{fig:setup}. Under these conditions
the elementary non-local transport process has been termed the quartet process. It corresponds 
to the exchange of two Cooper pairs, four quasiparticles, between the three superconductors, as
shown in Fig. \ref{fig:diagram_quartet}. 
The situation has been recently investigated in Refs. \onlinecite{Pfeffer}, \onlinecite{JeanEude}, and \onlinecite{Ronen}. 
Two of the first experiments study a diffusive TTJ, where
robust transport anomalies\cite{Pfeffer} and Shapiro steps\cite{JeanEude} 
were observed as a function of two applied voltages $V_{1,3}$, 
that have been interpreted in terms of three quartet modes. The third experiment studies a 
phase-coherent TTJ realized in a semiconducting nanowire\cite{Ronen}, showing positive current 
cross-correlation that are interpreted as evidence of the non-local quartet processes.

Motivated by these recent experiments, in this paper we describe the proximity effect in a
short, metallic TTJ under voltage bias, $V$ and $-V$, as shown in Fig. \ref{fig:setup}. We argue that driving
a dc current in terminal $2$, that is assumed at zero voltage, enables the control of the
static non-local phase governing the quartet process, $\varphi_Q$. We calculate the normal 
density of states (NDOS) in a wide parameter regime by employing the quantum circuit formulation 
of the quasiclassical Usadel equation \cite{Yuli,Yulibook,Vanevic}. For comparison we study the NDOS
in the biased two-terminal junction. We reveal the characteristic rich structure of the NDOS originating
from MAR, that is similar between two- and three-terminal junctions. We additionally reveal features
characteristic only to the three-terminal junction. The most striking of these are the pseudogaps appearing
in the NDOS in the regime where coherent non-local processes give rise to bound states. pseudogaps differ from
the proximity-induced minigap in that their edges are not as sharp, they do not in all regimes resemble the edges of the bulk gap, 
and may be less pronounced. What makes pseudogaps unique is the combination of properties: i. they are
tunable by the quartet phase, and ii. they depend strongly on voltage bias.

Our study includes the importance of electron-hole decoherence, introduced phenomenologically using
the quasiparticle dwell time in the normal region, $\tau_d$. Despite describing a short junction on the scale
of the coherence length, the dwell time can become appreciable compared to $\hbar/\Delta$ if the contact 
resistance at the $SN$ interfaces, $R_b$, is much larger than the intrinsic resistance of the junction $G_N^{-1}$. 
The Thouless energy \cite{Thouless} is proportional to the inverse dwell time and can be decreased by a factor $R_bG_N\gg 1$.
For this reason the Thouless energy can become comparable to or smaller than the superconducting gap, $\Delta$, 
even in short junctions. The magnitude of the proximity-induced minigap 
in the NDOS is drastically modified by decoherence effects in a large variety of Josephson junctions 
\cite{Dubos,proximity_expt,Gueron,minigap}.

We begin our presentation in Section \ref{phenomenology} with a phenomenological description of dynamics in a TTJ under voltage bias. 
The theoretical method and equations of quasiclassical circuit theory are presented in Section \ref{microscopic}. Section \ref{twoterminals} discusses the NDOS of a voltage-biased two-terminal Josephson junction, with peaks interpreted in terms of MAR processes. Section \ref{biasedTTJ} discusses the NDOS of a biased TTJ, revealing the signature of MAR processes as well as pseudogaps originating from non-local processes. Section \ref{conclusions} presents our conclusions. 

\section{Phenomenological description}
\label{phenomenology}

\subsection{Local and non-local Josephson effect}

The Josephson effect in an $N$-terminal Josephson junction is governed by $N-1$ independent superconducting phase differences.
Due to $2\pi$-periodicity, the phase-dependent part of the junction energy can be expanded in harmonics.
For $N=3$ we choose the gauge $\varphi_2=0$ and express $E_J$ as a Fourier series in $\varphi_1$ and $\varphi_3$,
\begin{align}
 E_J(\varphi_1,\varphi_3) = \displaystyle\sum_{m_1,m_3} E_{(m_1,m_3)} e^{i (m_1 \varphi_1 + m_3 \varphi_3)}, 
\end{align}
where $m_1$ and $m_3$ are integers running along the entire real axis, 
and the Fourier coefficients $E_{(m_1,m_3)}$ are generally complex energies chosen such that $E_J$ is real.

We explore non-local transport by choosing to evaluate the current 
flowing from terminal $2$ into terminals $j=\{1,3\}$, given by 
$I_{j} = (2e/\hbar)\partial E_J(\varphi_1,\varphi_3)/\partial \varphi_{j}$, 
\begin{align}
 I_j(\varphi_1,\varphi_3) = \displaystyle\sum_{m_1,m_3} I_{j,(m_1,m_3)} e^{i (m_1 \varphi_1 + m_3 \varphi_3)}, 
\end{align}
where $I_{j,(m_1,m_3)}=(2e/\hbar) im_jE_{(m_1,m_3)}$.
The total current flowing into terminal 2 is obtained from current conservation, $I_1+I_2+I_3=0$.
Any possible current flowing from terminal $1$ into terminal $3$ does not modify the discussion.

We define the non-local component of the current flowing from terminal $2$ into terminals $j=\{1,3\}$ 
by $I_{\rm j,NL}=\partial^2 I_j/\partial \varphi_1 \partial\varphi_3$.
The harmonic structure of the Josephson current permits identification of local terms, 
giving $I_{\rm j,NL}=0$, and non-local terms, giving rise to a finite $I_{\rm j,NL}$.
Three contributions correspond to the local Josephson effect between terminals: 
$1$ and $2$ given by harmonics $(m_1,0)$; $2$ and $3$ given by harmonics $(0,m_3)$, 
and $1$ and $3$ given by harmonics $(-m,m)$, with $-m_1=m_3=m$.
All other pairs of harmonics correspond to the non-local Josephson effect.

\begin{figure}
\begin{tabular}{|c|}
	\includegraphics[width=.6\linewidth]{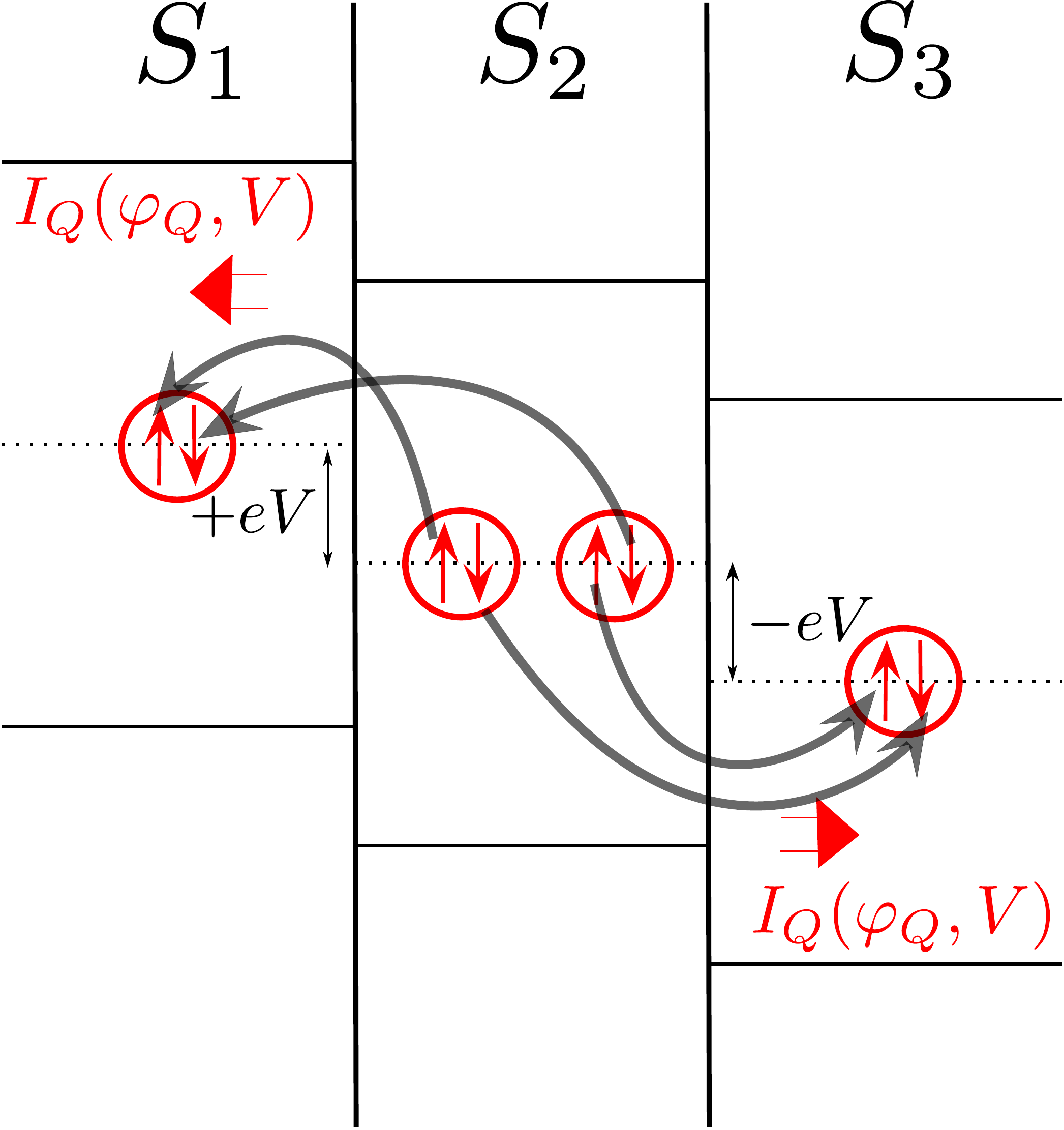}
\end{tabular}
\caption{(Color online.) Diagram of the four particle quartet process, i.e. the lowest order non-dissipative 
transport process. The resulting current, $I_Q$, is a dc current sensitive to voltage, $V$, as well
as the quartet phase, $\varphi_Q=(\varphi_1-\varphi_2)+(\varphi_3-\varphi_2)$.}
\label{fig:diagram_quartet}
\end{figure}

The non-local Josephson term lowest in the order of harmonics corresponds to $(m_1,m_3)=(1,1)$. 
It has been named the quartet term, as it implies a coherent exchange of two Cooper pairs, 
four quasiparticles, between the superconductors as shown in Fig.~\ref{fig:diagram_quartet}.
In the following we show how the quartet term can be filtered from terms corresponding 
to the rest of the harmonics when driving the junction under commensurate voltage bias, 
$V_1=-V_3=V$.

\subsection{Out-of-equilibrium dynamics}

Under commensurate voltage bias, $V_1=-V_3=V$, $V_2=0$, the phases are given by:
$\varphi_2=0$, $\varphi_1=\varphi_{10} + \omega t$, and $\varphi_3=\varphi_{30} - \omega t$, 
where $\omega=2eV/\hbar$ is the Josephson frequency. The effect of biasing is to
separate the harmonics of the Josephson energy in frequency space.

Under these biasing conditions, the quartet term and its harmonics $(m,m)$
give rise to dc current in terminal $j=\{1,3\}$,
\begin{align}
 I_{\rm j,DC}(\varphi_Q,V) = I_Q = \displaystyle\sum_{m} I_{m}(V) \exp(i m \varphi_Q ), 
\end{align}
where $\varphi_Q=\varphi_{10}+\varphi_{30}$ is the quartet phase and $I_m=(2e/\hbar)imE_{(m,m)}$.
A detailed discussion of the coefficients $I_m$ and their dependence on the bias voltage will be
presented elsewhere.



The quartet phase can be tuned independently of the bias voltage by imposing an external current in
terminal $2$. Current conservation leads to a current-phase dependence, $I_{DC}(\varphi_Q)$, similar to the
dc Josephson effect, $I_{DC} = -I_2 = 2I_Q(V, \varphi_Q)$. The indirect control
of the quartet phase by current bias is similar to the control of the phase drop in a two-terminal 
Josephson junction by dc current bias. In analogy, the dc current is $2\pi$-periodic in the quartet phase.
If $I_{DC}$ surpasses a certain critical value, depending on the details of the junction, the dc
behavior of the junction becomes resistive. This situation, together with a discussion of the 
current flowing between terminals $1$ and $3$, will be presented in detail elsewhere.
For discussing the proximity-induced normal density of states (NDOS) in the junction, 
we will use $V$ and $\varphi_Q$ as independent control parameters. 

\section {Microscopic model}
\label{microscopic}

We describe transport in a metallic TTJ using quasiclassical equations of non-equilibrium superconductivity. These take the form of a diffusive equation for the quasiclassical Keldysh-Nambu Green's function, \cite{Larkin} also known as the Usadel equation (see also Ref. \onlinecite{Yulibook})
\begin{align}
\ & \frac{\partial}{\partial \textbf{x}}\left({\cal D}(\textbf{x})\check{G}\frac{\partial}{\partial \textbf{x}}\check{G}\right)-i\left[\check{H},\check{G}\right]=0,\\
\ & \check{G} = \Mab{G^R}{G^K}{0}{G^A}\ , \quad \check{G}^2=\check{1},\quad \check{H} = \Mab{\hat{H}}{0}{0}{\hat{H}}; \notag\\ 
\ & \hat{H}=E \hat{\sigma}_z+\frac{1}{2}\Delta(\textbf{x})(i\hat{\sigma}_y+\hat{\sigma}_x)+\frac{1}{2}\Delta^*(\textbf{x})(i\hat{\sigma}_y-\hat{\sigma}_x). \notag
\end{align}
In addition to Keldysh-Nambu space (denoted with a check hat, $\check{G}$), the quasiclassical Green's function generally depends on two times (or energies $\bf E$) and on spatial coordinates $\check{G}(\bf E,\textbf{x})$. The Pauli matrices are defined in Nambu space (denoted with a hat) $\vec{\hat{\sigma}}={\hat{\sigma}_x, \hat{\sigma}_y, \hat{\sigma}_z}$, and ${\cal D}(\textbf{x})$ denotes the diffusion coefficient. Matrix products in the Usadel equation are understood as convolutions of the quantities in the double time (or energy) representation, as detailed in the Appendix.

The Usadel equation applies to the most common experimental situation where the junction dimensions are larger than the elastic mean-free path. It is a conservation equation for the Keldysh-Nambu current density, $\check{j}(\textbf{x})$,
\begin{align}
\label{usadelcurrent}
\frac{\partial}{\partial \textbf{x}}\check{j}(\textbf{x})+\frac{ie^2\nu}{\hbar}\left[\check{H},\check{G}\right]=0;\quad \check{j}=-\sigma(\textbf{x})\check{G}\frac{\partial}{\partial \textbf{x}}\check{G}.
\end{align}
Here, $\nu$ is the electronic NDOS and $\sigma(\textbf{x})$ is the conductivity. The two quantities are related by $\sigma = e^2{\cal D}\nu$.

Hereafter we employ a discretized version of the Usadel equation that describes the system in terms of finite quantum circuit elements \cite{Yuli,Yulibook}. The bulk superconducting terminals $S_i$ are described by coordinate-independent Keldysh-Nambu Green's functions $\check{G}_i$. The junction area is represented by a single node described by the unknown Green's function $\check{G}_c$. The node is separated from each terminal $S_i$ by a connector that models the transparency of the contact via transmission coefficients $T^{(i)}_n$ corresponding to channel $n$ in contact $i$. The Keldysh-Nambu matrix current flowing between terminal $S_i$ and the node takes the compact form,\cite{Yuli}
\begin{equation}
I_{ic} = \frac{2e^2}{\pi\hbar}\sum_n \frac{ T^{(i)}_n [\check{G}_i,\check{G}_c]}{4+T^{(i)}_n(\{\check{G}_i,\check{G}_c\}-2)}.
\label{eq:currentformula}
\end{equation}
The fraction notation for matrix inversion is justified since $\check{G}_i$ and $\check{G}_c$ commute with $\{\check{G}_i,\check{G}_c\}$.

Decoherence between electrons and holes is accounted for phenomenologically 
by connecting the node to a fictitious terminal \cite{Yuli}. In contrast to the other three terminals, that correspond to the superconductors, the Keldysh-Nambu current flowing in the fictitious terminal does not contain particle or energy currents. The Green's function of the fictitious terminal is chosen such that the corresponding Keldysh-Nambu current describes only the leakage of electron-hole coherence. The Keldysh-Nambu current matrix to the fictitious terminal is given by,
\begin{align}
I_{fc} =& \frac{2e^2}{\pi\hbar}\sum_i \sum_n \frac{T_n^{(i)}}{4} [\check{G}_f,\check{G}_c]\ ,\label{eq:fictitious}\\
\check{G}_f =&  -i \frac{E\tau_d}{\hbar} \Mab{\hat{\sigma}_z}{0}{0}{\hat{\sigma}_z}, \notag
\end{align}
where $\tau_d$ is the dwell time of quasiparticles in the junction, including the connectors. 
By including $I_{fc}$, the transport equation can be written as a conservation of the current of coherences,
\begin{equation}
\sum_i \check{I}_{ic} + \check{I}_{fc}= 0\ .
\end{equation}
Since each of the currents are given by a commutation relation between the unknown Green's function of the central node $\check{G}_c$ and a matrix defined by Eqs.~(\ref{eq:currentformula}) and (\ref{eq:fictitious}), it is convenient to rewrite the current conservation as a commutation relation $[\check{G}_c,\check{M}] = 0$, where the matrix denoted by $\check{M}$ adds up the terms corresponding to the four currents,
\begin{equation}
\check{M} = \displaystyle{\sum_{i,n}}\ T_n^{(i)}\left(\frac{\check{G}_i}{1+\frac{T_n^{(i)}}{4}(\{\check{G}_i,\check{G}_c\}-2)}+\check{G}_f\right).
\label{currentconservation}
\end{equation}
It is important to note that matrix $\check{M}$ depends non-linearly on the unknown Green's function of the central node $\check{G}_c$, as well as on the known Green's functions of the terminals. The relation $[\check{G}_c,\check{M}(\check{G}_c)] = 0$ is a non-linear equation to be solved numerically for $\check{G}_c$.

\subsection{Green's functions of superconducting terminals}

In equilibrium, transport is stationary and the Green's functions depend on a single energy (or, in time representation, on the difference of the two times and independent of their sum). As a function of energy, the Green's functions of the superconducting terminals are given by,
\begin{align}
\label{fgeq}
{G}^R_i = \frac{1}{\xi}\Mab{\epsilon}{\Delta_i}{-\Delta_i^*}{-\epsilon}\ ;\quad {G}^A_i = -\frac{1}{\xi^*}\Mab{\epsilon^*}{\Delta_i}{-\Delta_i^*}{-\epsilon^*}\ ,
\end{align}
where complex energies have been introduced $\epsilon=E+i0^+$ and $\xi=\sqrt{\epsilon+|\Delta|}\sqrt{\epsilon-|\Delta|}$. Here, $\Delta_i=|\Delta|e^{i\varphi_i}$. The positive, vanishing imaginary part of $\epsilon$ specifies the position with respect to the branch cut of the square root function in the complex plane.

The advanced and retarded Green's functions are related by $G^A = -\hat{\sigma}_z \left(G^{R}\right)^{\dagger} \hat{\sigma}_z$ and the Keldysh Green's function $G^K$ is obtained from:
\begin{equation} 
G^K=(G^R - G^A)\tanh(\beta E/2),
\end{equation}
where $\beta=(k_BT_e)^{-1}$ ($T_e$ is the temperature).

We consider voltage-biased terminals, $V_1=-V_3=V,V_2=0$. According to the second Josephson relation, $\dot{\varphi}_i=2eV_i/\hbar$, constant voltage bias gives rise to time-dependent superconducting phase differences that in general give rise to non-stationary transport. As a result, Green's functions acquire a non-trivial dependence on both energies, or equivalently in time representation, on both the difference, $(t_1-t_2)$, as well as the sum, $(t_1+t_2)$ of the two times. We relate the out-of-equilibrium Green's function of terminal $i$ to its equilibrium value by the following gauge transformation,
\begin{equation}
\label{gaugetransfo}
\check{G_i}(t_1,t_2)=e^{i\hat{\sigma}_zeV_it_1}\check{G}_i(t_1-t_2)e^{-i\hat{\sigma}_zeV_it_2}.
\end{equation}

\subsection{Numerical implementation}

The theoretical framework outlined so far is sufficiently general to describe out-of-equilibrium transport for arbitrary bias. However, the non-linear equations that determine the unknown Green's function of the node, $\check{G}_c(t_1,t_2)$, are very difficult to solve in general. The dependence on two times (or two energies) must be solved on a discrete grid, where each grid point corresponds to an entry of the unknown matrix $\check{G}_c(t_1,t_2)$ (keeping in mind that each entry is a $4\times 4$ matrix in Keldysh-Nambu space). In the general case the size of matrices involved grows quickly giving rise to an overwhelming computational problem. 

To proceed, we use the properties of commensurate bias. In general, transport is governed by two Josephson frequencies corresponding to the two independent voltage differences. For commensurate bias, the two Josephson frequencies are harmonics of a single frequency $\omega_0$, the greatest common divisor. For the specific bias $V_1=-V_3=V,V_2=0$, the greatest common divisor is the Josephson frequency $\omega_0=2eV/\hbar$.
We take advantage of this property by performing a double-time Fourier transform, (detailed in the Appendix) previously used in a different context in Ref. \onlinecite{Jonckheere2009}. In the transformed representation the Green's functions depend on a single energy (as in equilibrium) and on the harmonics of $\omega_0$ counted by two indices, $\check{G}(E,n,m)$. 
The definition contains redundancy in the indices, $\check{G}(E,n,m)=\check{G}(E-p\omega_0,n+p,m+p)$, therefore the Green's functions are determined by the value in the energy interval $[-\omega_0/2,\omega_0/2]$. (here we have set $\hbar=1$)
An alternative representation with only one harmonic index 
has been used in Ref. \onlinecite{Bezuglyi2} for a two-terminal Josephson junction in the tunnel limit.


\begin{figure}[t!]
\begin{center}
\includegraphics[width=9pc]{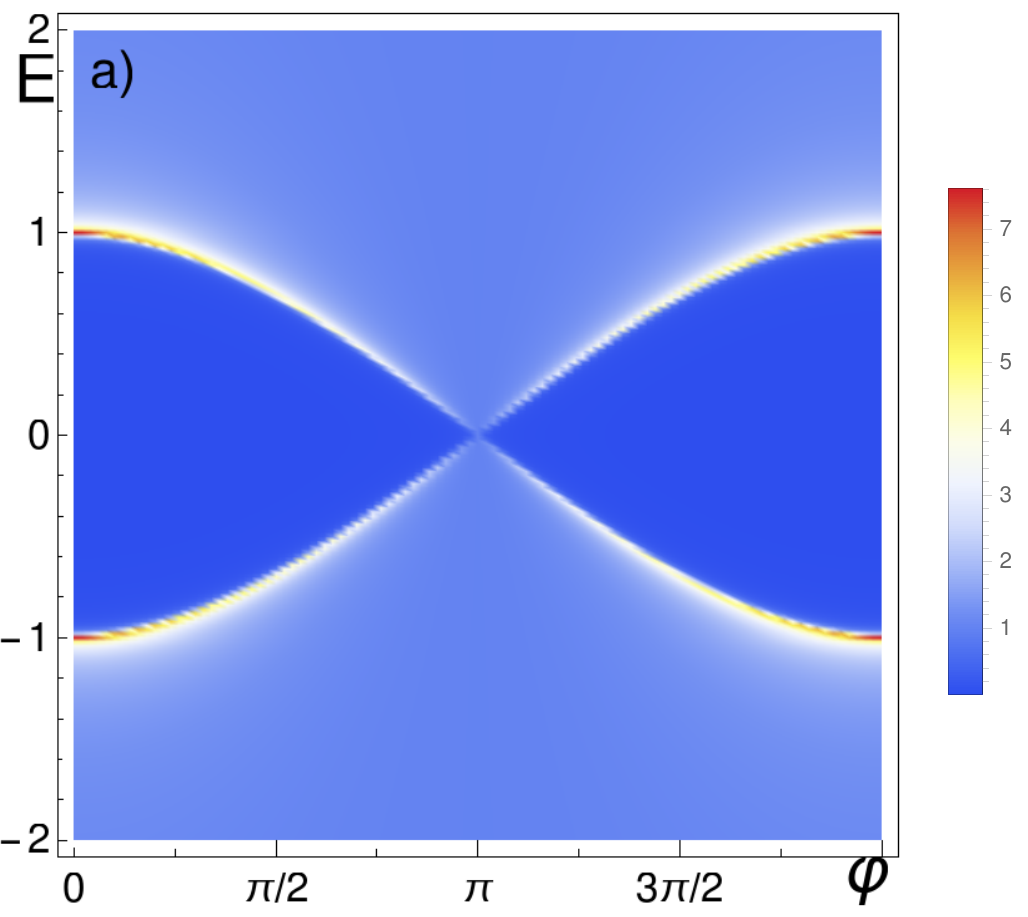} \hspace{0.1cm}
\includegraphics[width=9pc]{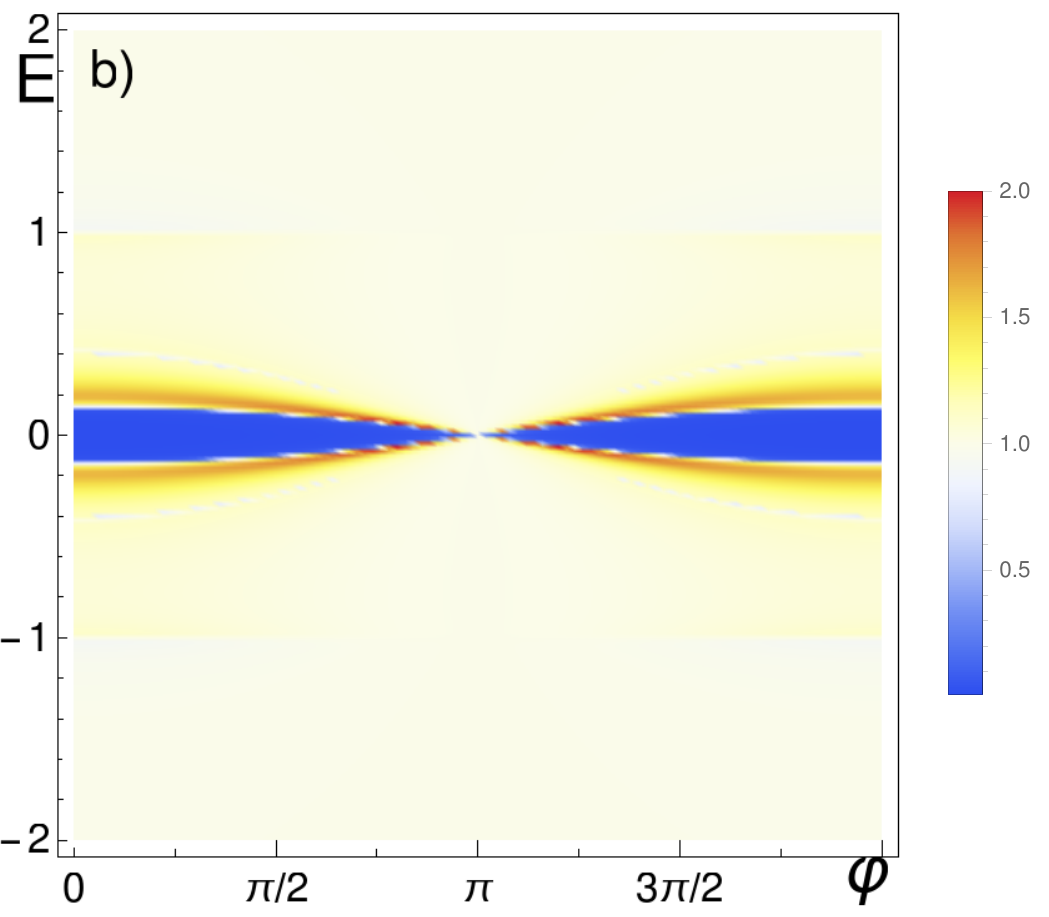} \vspace{0.3cm} \\ 
\includegraphics[width=9pc]{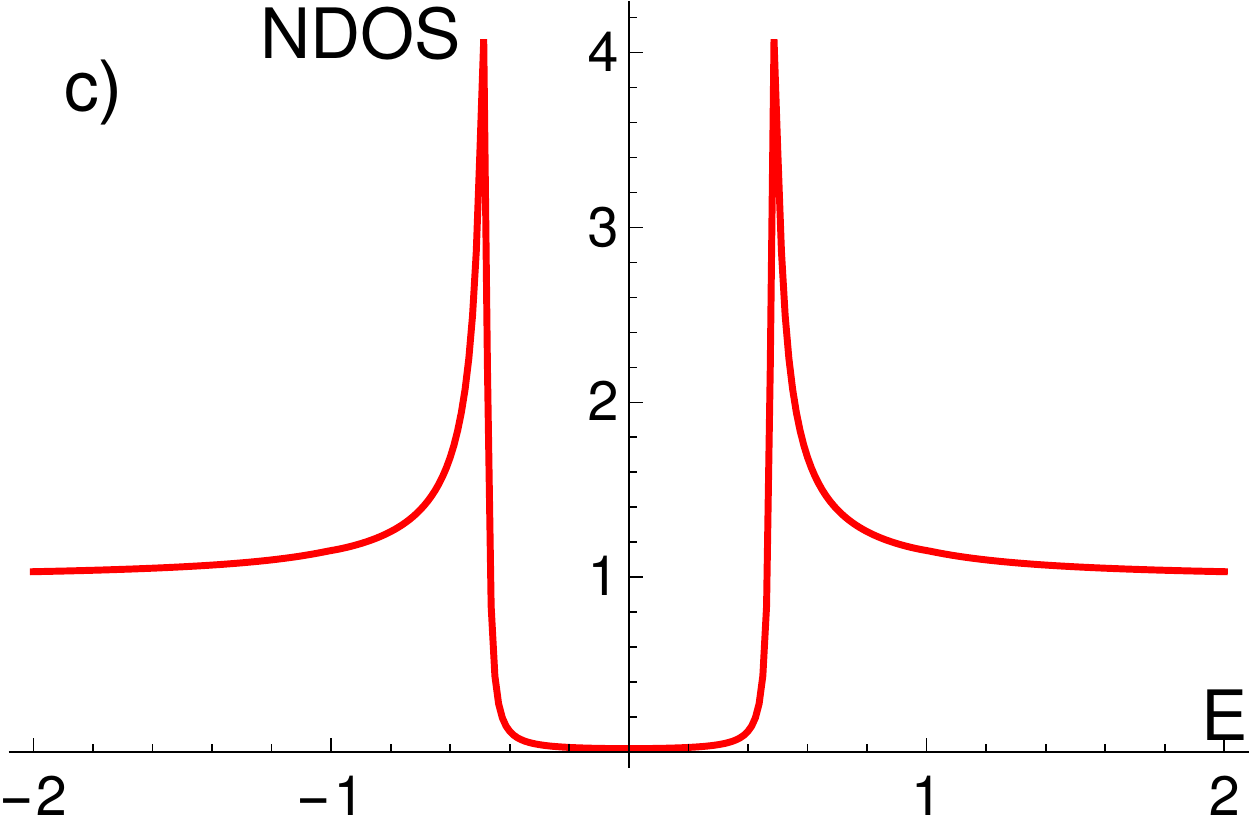}\hspace{0.1cm}
\includegraphics[width=9pc]{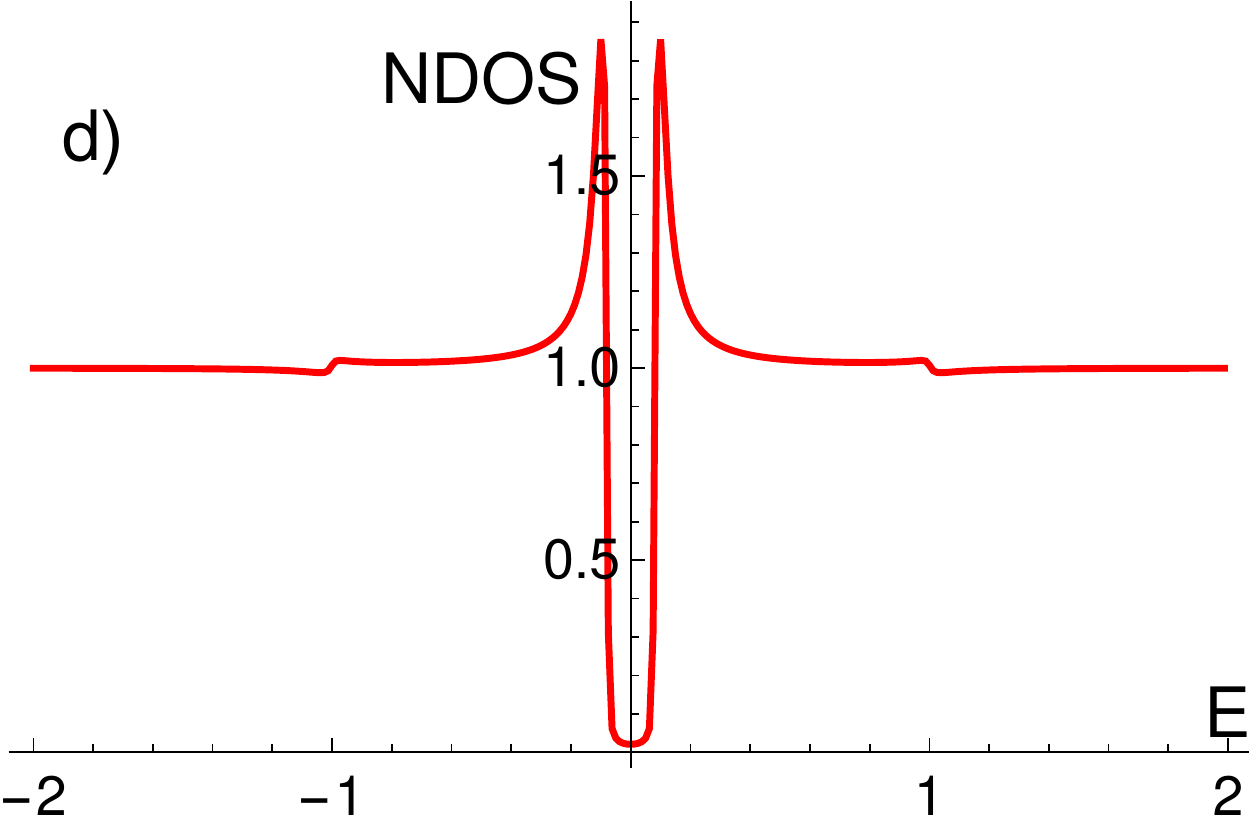} \vspace{0.3cm}
\caption{(Color online.) Normal density of states (NDOS) for a
symmetric SNS junction ($T=0.3$), for the small decoherence case (left column, $\tau_d=0.05$),
and for the large decoherence case (right column, $\tau_d=5.0$).  
(a) NDOS as a function of the phase $\varphi$ and of the 
energy $E$, for $\tau_d=0.05$. 
(b) Same as (a), with $\tau_d=5.0$. (c) Cut of (a) for the phase $\varphi=2\pi/3$.
(d) Cut of (b) for the phase $\varphi=2\pi/3$. Energy $E$ is measured in units of $\Delta$
and $\tau_d$ in units of $\hbar/\Delta$. ($\Delta=\hbar=e=1$.)}
\label{DOSeqJJ}
\end{center}
\end{figure} 

\begin{figure*}[t]
 \centerline{\includegraphics[width=14pc]{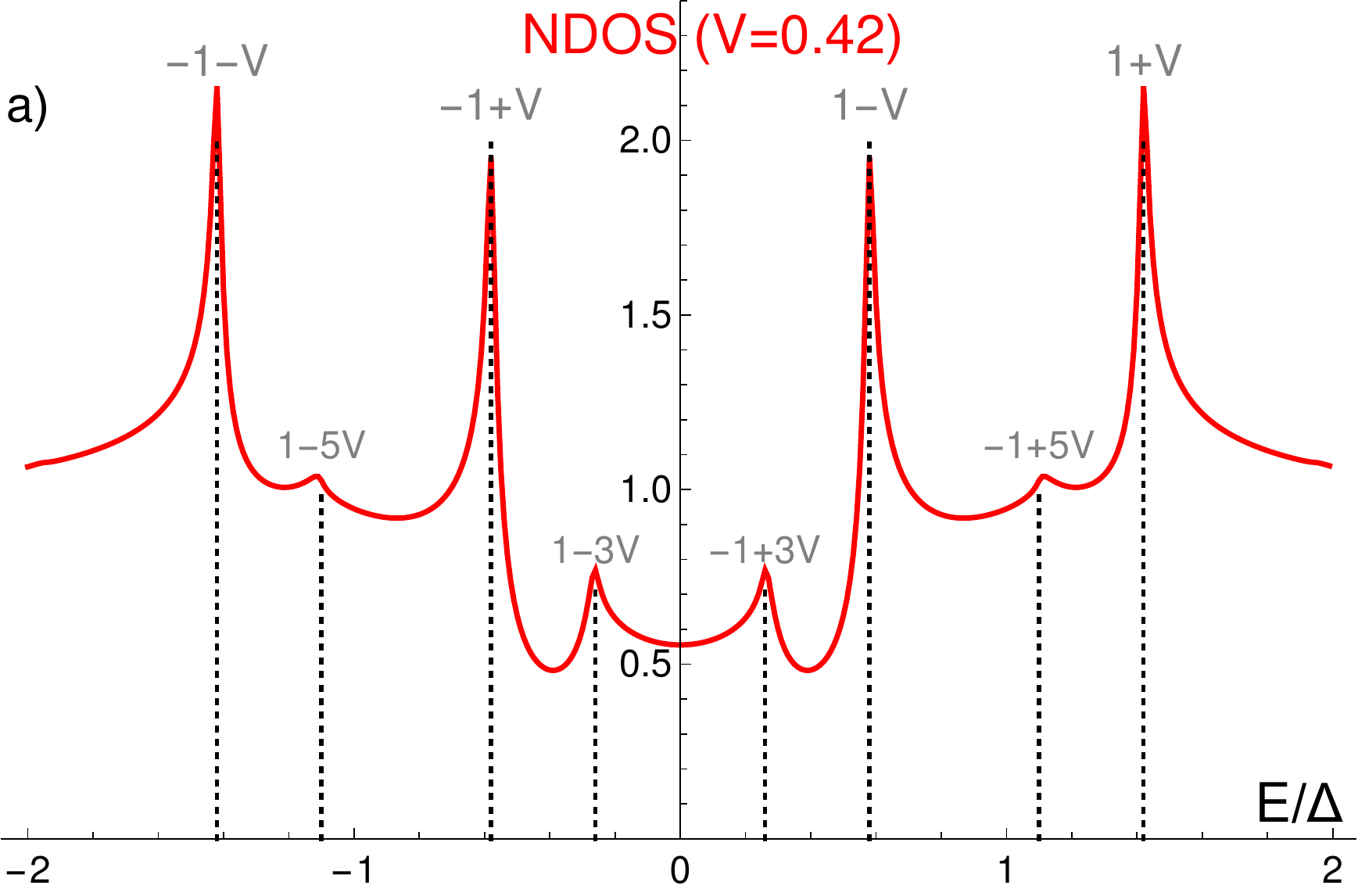} \hspace{0.1cm}
 \includegraphics[width=14pc]{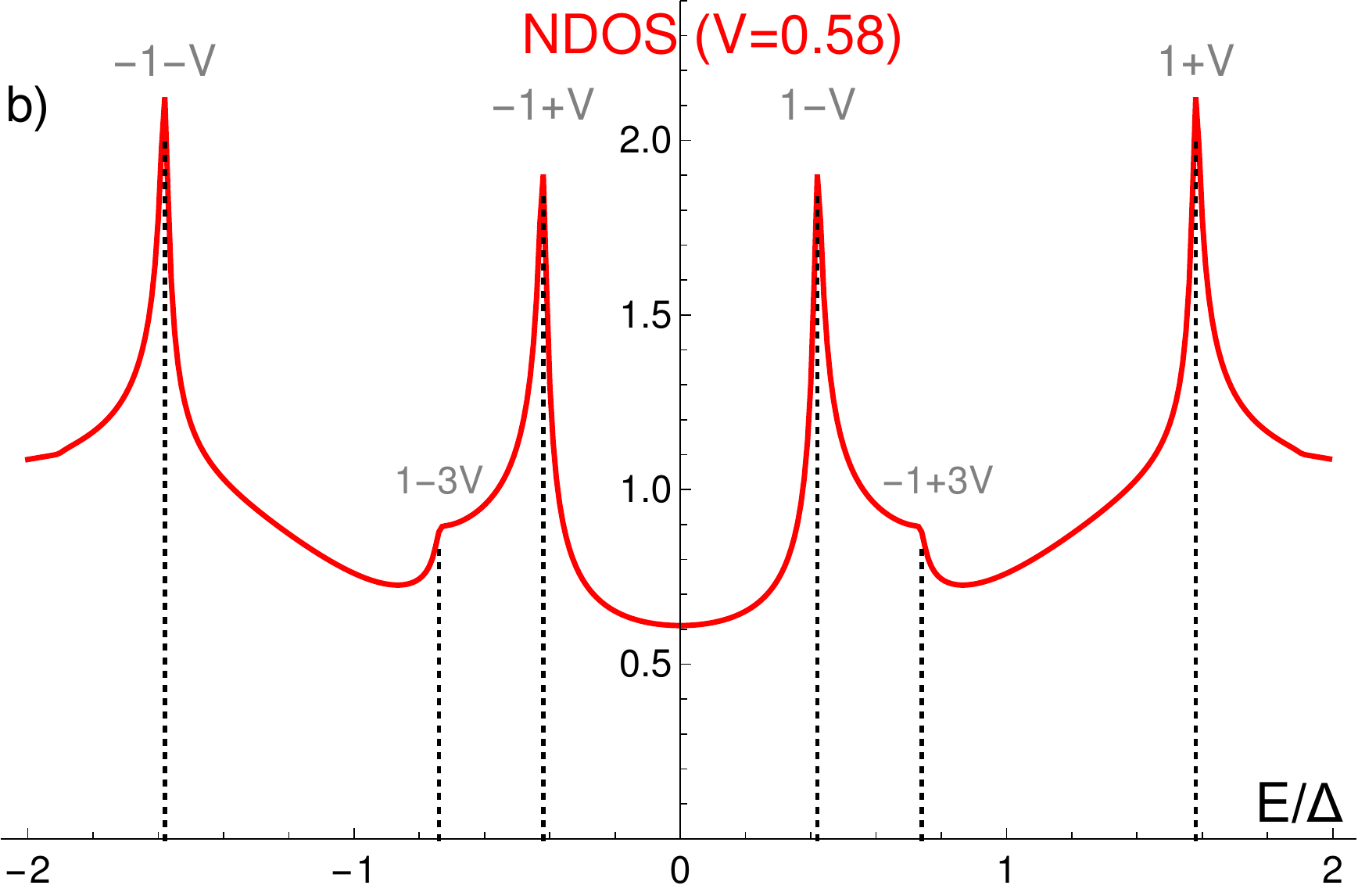}\hspace{0.2cm}
 \includegraphics[width=10pc]{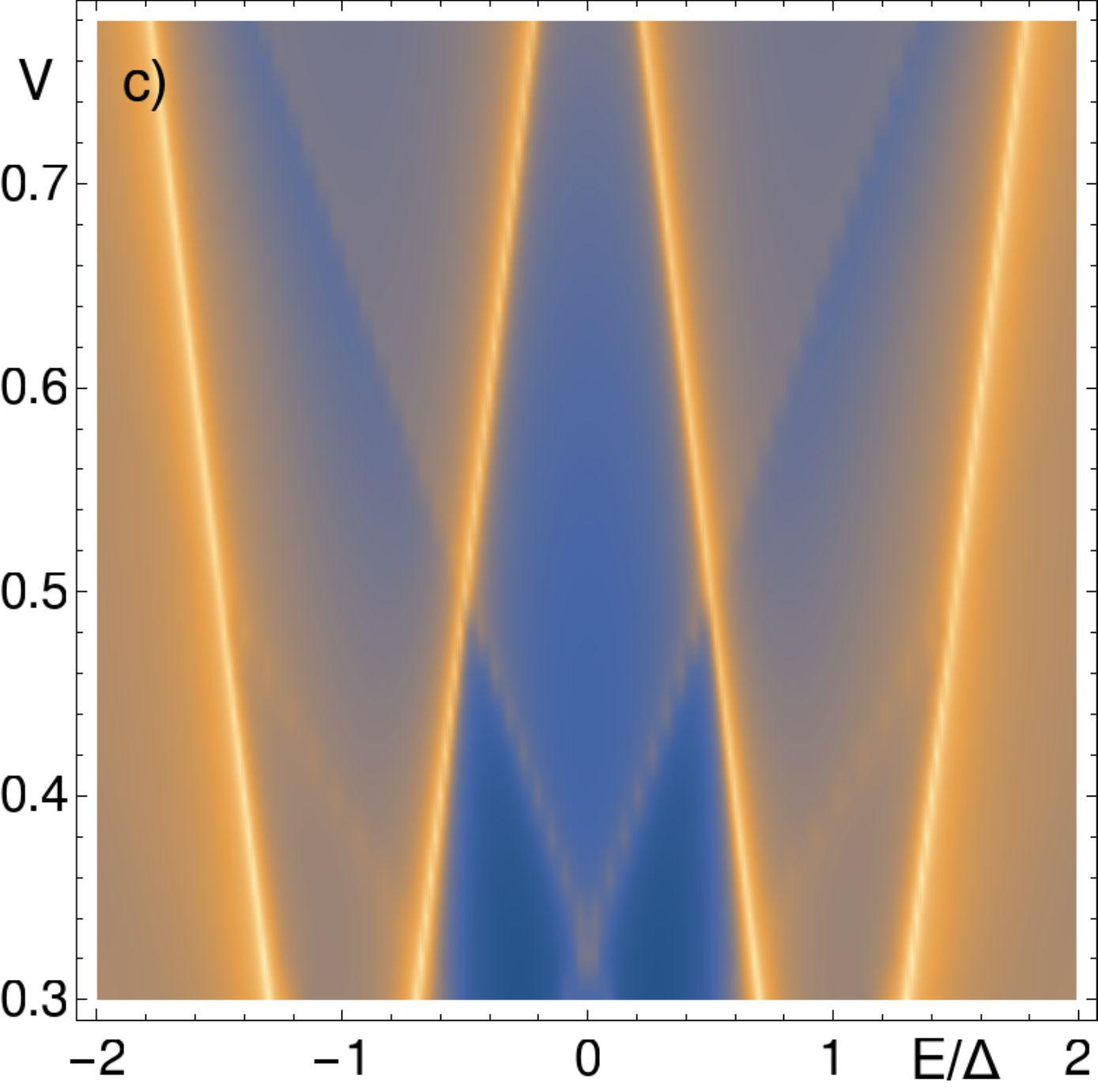}}
 \caption{(Color online.) Normal density of states (NDOS) as a function of energy $E$ in a biased junction at transparency $T=0.1$,
  for small decoherence ($\tau_d=0.05$). (a) and (b) show the NDOS for $V=0.42$ and 
  $V=0.58$, and (c) shows the density plot of the NDOS as a function of the energy $E$ and the voltage $V$. In (a) and (b), the vertical dotted lines show the positions of the expected peaks at voltages $\pm(\Delta\pm(2p+1)V)$ due to MAR processes. ($\Delta=\hbar=e=1$.)}
 \label{fig:DOSbiasedJJ1smalltd}
\end{figure*}

\begin{figure*}[t]
 \centerline{\includegraphics[width=14pc]{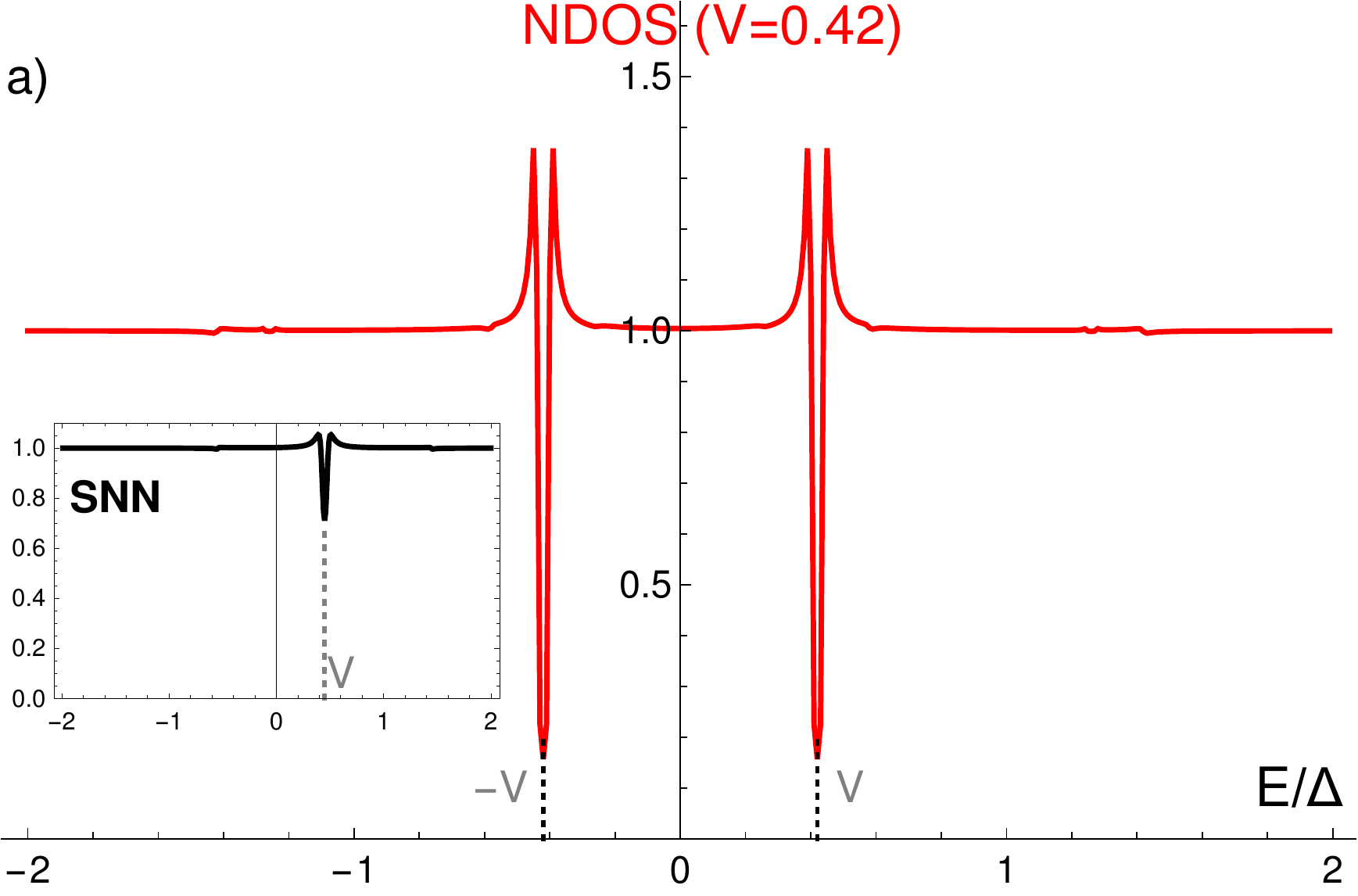} \hspace{0.1cm}
 \includegraphics[width=14pc]{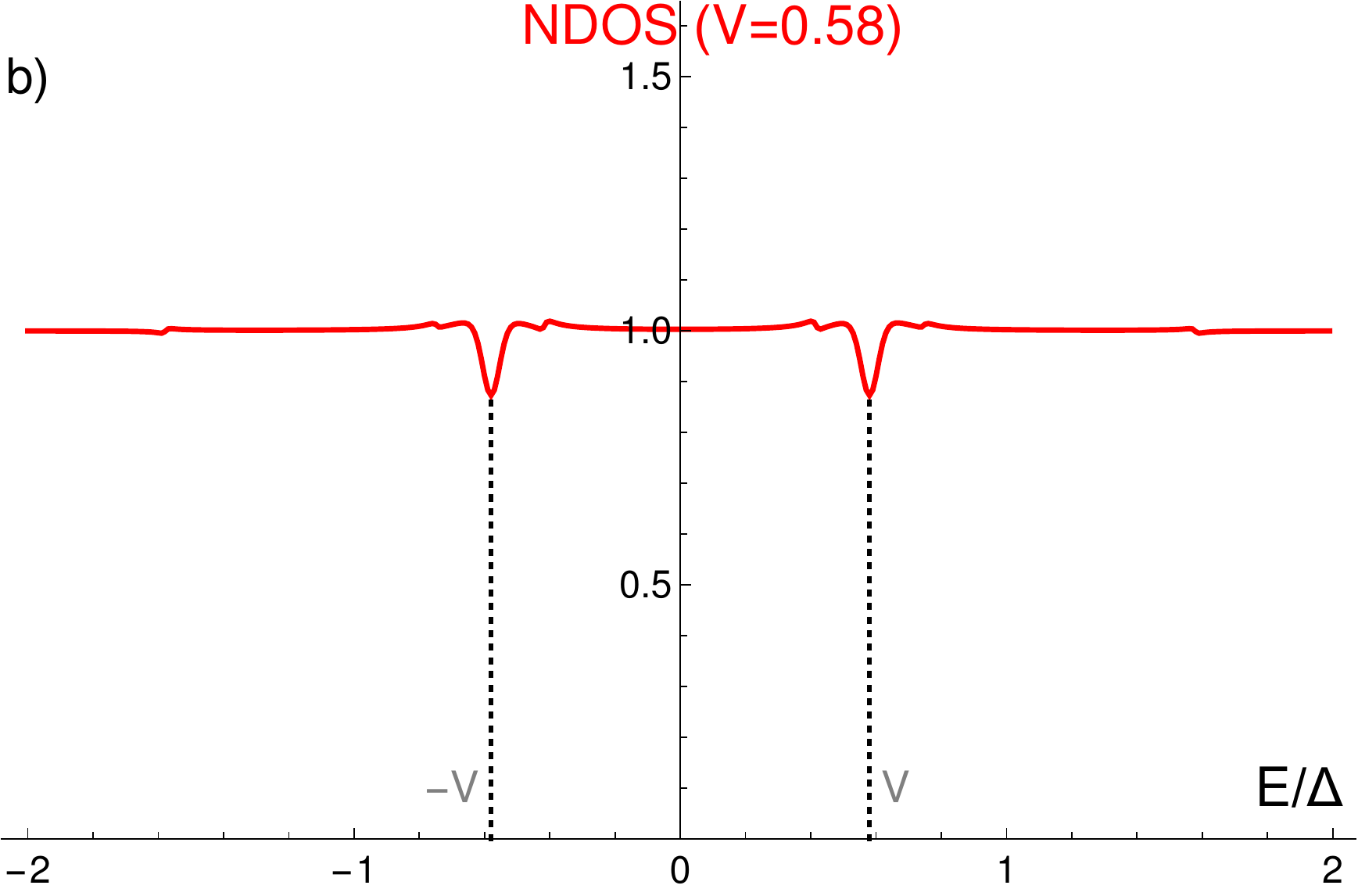}\hspace{0.2cm}
 \includegraphics[width=10pc]{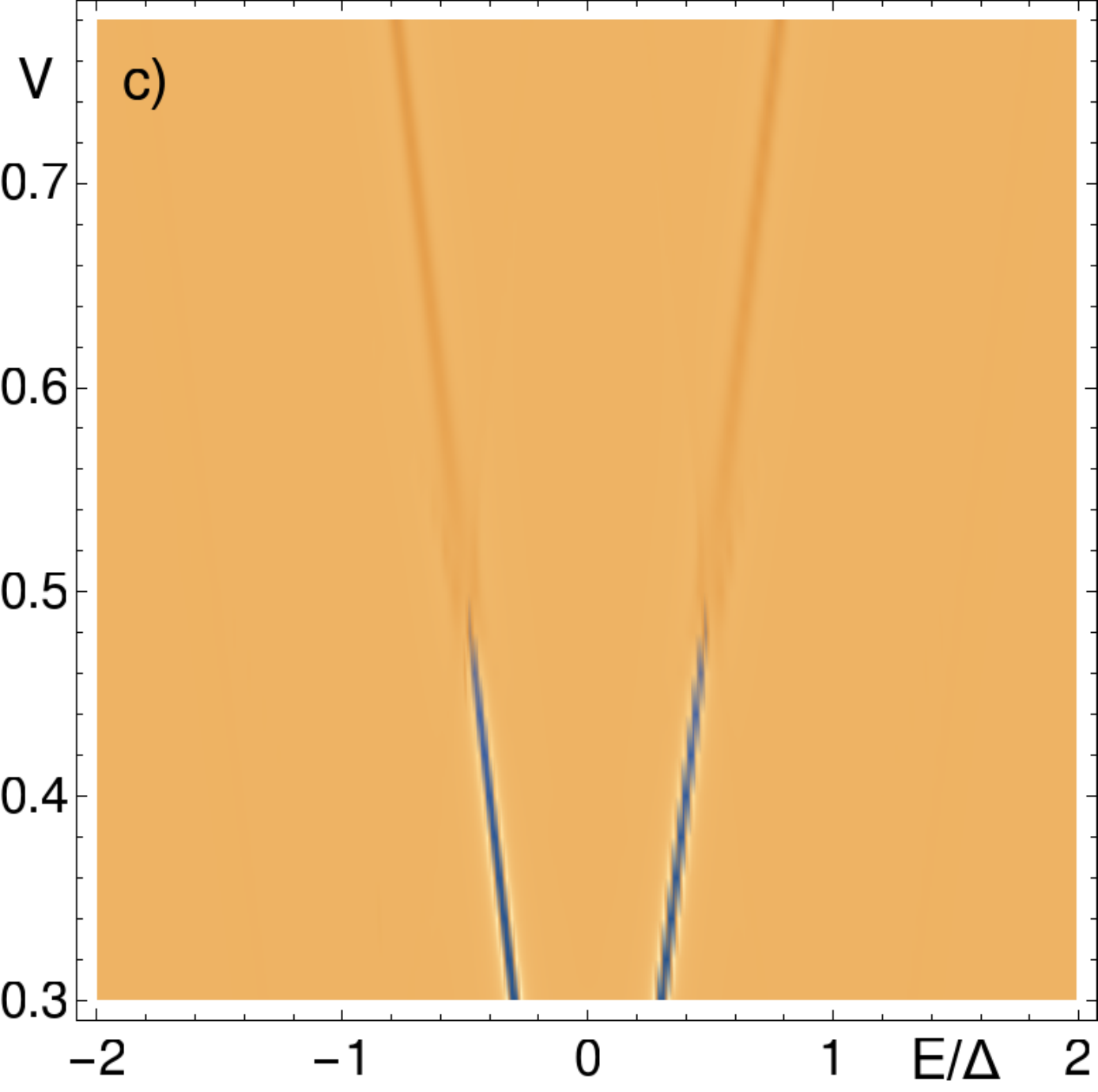}}
 \caption{(Color online.) Same as Fig.\ref{fig:DOSbiasedJJ1smalltd}, but for large decoherence ($\tau_d=5.0$).
 In (a) and (b), the vertical dotted lines show the positions of the expected minigaps peaks at voltages $\pm V$ corresponding to the chemical potentials of the two electrodes. The
 inset in (a) shows the NDOS computed for a SNN junction were the $S$ electrode is biased at voltage $V=0.45$, with $\tau_d=5$.  }
 \label{fig:DOSbiasedJJ1largetd}
\end{figure*}
 
The practical numerical implementation involves the truncation of the harmonics by a value $N_m$, whereby the Green's functions are square matrices of dimension $4(2N_m+1)$ defined on a one-dimensional grid in the energy interval $[-\omega_0/2,\omega_0/2]$. The matrix entries decay quickly at large harmonics. It is sufficient to truncate the harmonic expansion at $N_m=(2\Delta/eV)$. 

We take the following steps to solve for the unknown Green's function of the central node, $\check{G}_c$, iteratively at each energy: (i) we start with a guess value for $\check{G}_{c,n}$, (ii) obtain and diagonalize $\check{M}_n(\check{G}_{c,n})$, (iii) obtain a new value, $\check{G}_{c,\text{new}}$ that commutes with $\check{M}_n$ and has as eigenvalues the signs of the real part of the eigenvalues of $\check{M}_n$, (iv) we check if $\check{G}_{c,\text{new}}$ is within a certain tolerance of $\check{G}_{c,n}$, to ascertain convergence, and finally, (v) if convergence was not achieved, we use a modified matrix as the guess of the next iteration step, $\check{G}_{c,n+1}=||\check{G}_{c,n}+\alpha \check{G}_{c,\text{new}}||$, where $\alpha<1$ is a convergence parameter that must be reduced at energies where the transport depends sharply on energy, and $||...||$ denotes the normalization that ensures $\check{G}_{c,n+1}^2=1$. In the calculation, a finite imaginary part is added to the energy, $\varepsilon=E+i\eta$, with $\eta/\Delta\ll 1$, to generate numerically smooth transport resonances. The parameter $\eta$ may be understood as a phenomenological description of weak inelastic effects. Convergence to the $\eta=0+$ limit is especially slow for all superconducting multi-terminal calculations \cite{REGIS}, requiring small convergence parameters of the order $\alpha\simeq \eta/\Delta$. In the numerical calculation we have used $\eta=0.01$.

\subsection{The density of states (NDOS)}

The NDOS can be measured using a tunnel probe, as
has been already realized for a three terminal junction in equilibrium \cite{Giazotto}.  
We model the tunnel probe by adding a normal terminal tunnel coupled to the junction.
The current to the tunnel probe is given by,
\begin{equation}
I_{t} = \frac{e^2}{2\pi\hbar} T_\text{tun} \Tr \Big(\hat{\sigma}_z[\check{G}_t,\check{G}_c]^K\Big),
\label{eq:tunnelcurrent}
\end{equation}
where $K$ denotes the Keldysh part of the matrix and $T_\text{tun}\ll 1$ describes the coefficient of the tunnel contact. 
Given that $G_t^{R,A}=\pm \hat{\sigma}_z$ and $G_t^K=4\tanh\big((\frac{\beta}{2}(\varepsilon \hat{\sigma}_z+V_t)\big)$, one can rewrite $I_t$ as:
\begin{equation}
I_{t} = \frac{e^2}{2\pi\hbar} T_\text{tun} \Tr\big[2 G_c^K+4\tanh\big(\frac{\beta}{2}(\varepsilon +\hat{\sigma}_z V_t)\big)(G_c^A-G_c^R)\big].
\label{eq:exptunnelcurrent}
\end{equation}
For small tunnel coefficients $T_\text{tun}$, 
the effect of the probe voltage $V_t$ on $\check{G}_c$ can be neglected. In this case, the tunnel conductance
, $\mathcal{G}_t=dI_t/dV_t$, is given by,
\begin{equation}
\mathcal{G}_t=\frac{8e^2}{\pi \hbar} T_\text{tun} \frac{d}{dV_t} \Tr\big[\tanh\big(\frac{\beta}{2}(\varepsilon +\hat{\sigma}_z V_t)\big)(G_c^A-G_c^R)\big].
\label{eq:tunnelconductance} 
\end{equation}
The expression of the tunnel conductance is identical to its expression in equilibrium, with the exception that here the term $(G_c^A-G_c^R)$ includes the dynamics driven by the bias $V$. We conclude that the out-of-equilibrium NDOS can be probed by tunneling spectroscopy, with the following observation. Care must be taken at energies where the NDOS presents sharp structures that result from the divergence of the bulk superconductor NDOS. At these energies the perturbative treatment of the tunnel probe may fail. For this reason tunnel measurement of the NDOS may give rise to a rounding of the sharpest features of the NDOS that we predict in absence of the probe.
In the following we will neglect the effect of the probe on transport and calculate the NDOS given by ${\cal N}(E)=\Re \check{G}_c^{11}(E,0,0)$. 

\section{The two-terminal junction}
\label{twoterminals}

\begin{figure}[h!]
\begin{center}
\includegraphics[width=20pc]{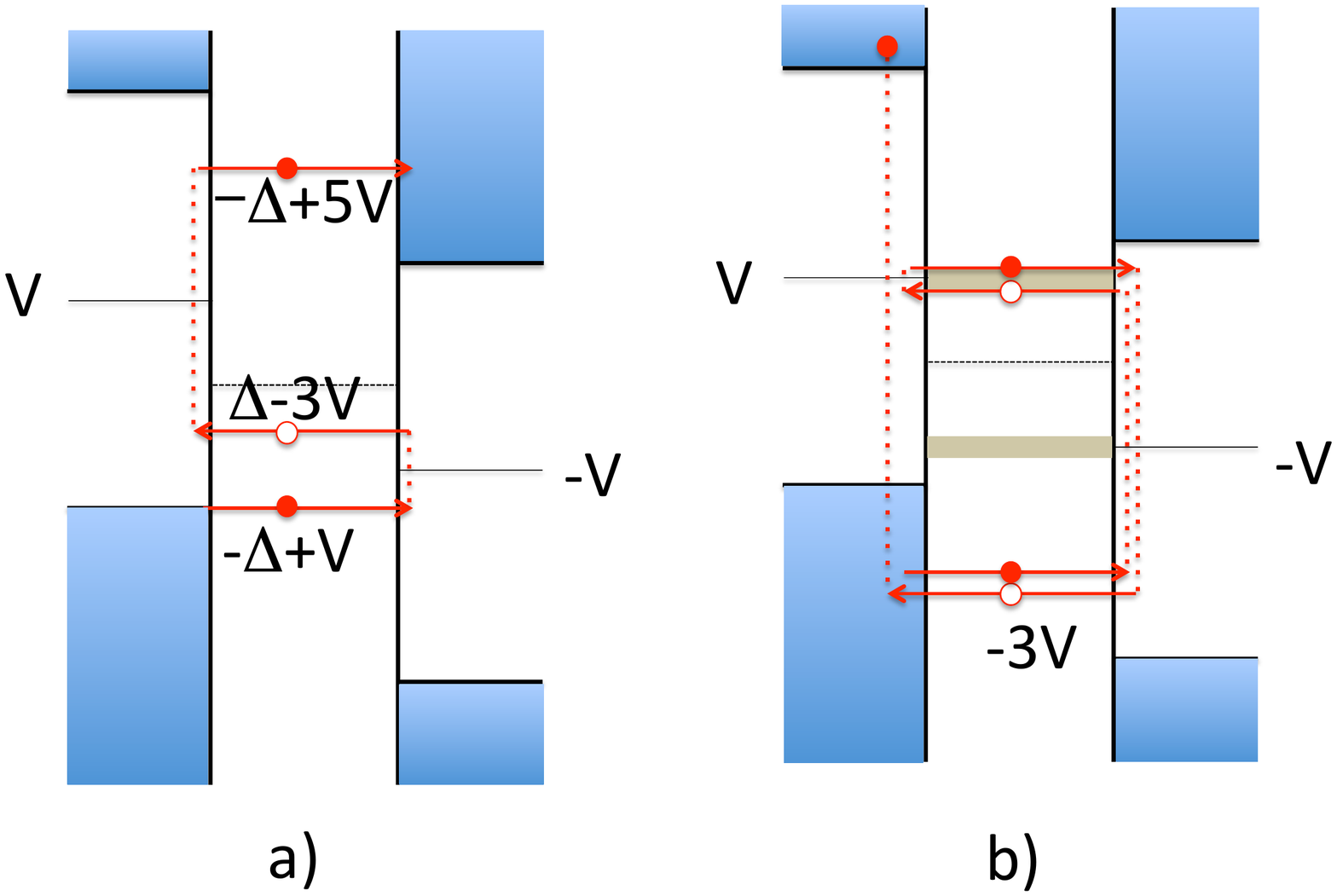}
\caption{(Color online.) Schematic diagrams showing the dominant MAR processes at lowest order in the number of Andreev reflections.  (a) No dephasing case ($n=3$), dominated by the gap edges ($eV=0.4\Delta$). (b) With dephasing ($n=4$), the (brown online) shaded areas denote the pseudogaps at $\pm eV$ ($eV=0.4\Delta$).
}
\label{figMARdiag}
\end{center}
\end{figure}

As a preliminary, we present the results of quantum circuit theory for the conventional, two-terminal Josephson junction, both in equilibrium as well as in a symmetric junction biased at ($V,-V$), for arbitrary transparency. This last problem has been considered by Bezuglyi et al., in the tunneling limit \cite{Bezuglyi,Bezuglyi2}. The choice of bias, $2V$, helps us make a direct comparison with the biasing conditions of the $N=3$ terminals junction.

\subsection{Equilibrium}
As a reference, the NDOS of a symmetric equilibrium SNS junction (with transparencies $T=0.3$) 
is represented in Fig.~\ref{DOSeqJJ}, for small ($\tau_d \ll 1$) and large  ($\tau_d > 1$) decoherence. 
(Throughout the paper we have set $\hbar=1$, the elementary charge $e=1$ and, unless explicitly shown, $\Delta=1$.
The dwell time $\tau_d$ is presented in units of $\hbar/\Delta$.)
The NDOS vanishes within the well-known minigap. The minigap persists for all parameters, except at 
$\varphi=\pi$ for the symmetric case, where the resonance gives rise to perfect transmission.  
For strong decoherence, the NDOS exhibits a sharp minigap that scales as the Thouless energy.


\subsection{Biased junction}

\begin{figure}[h!]
\begin{center}
 \includegraphics[width=14pc]{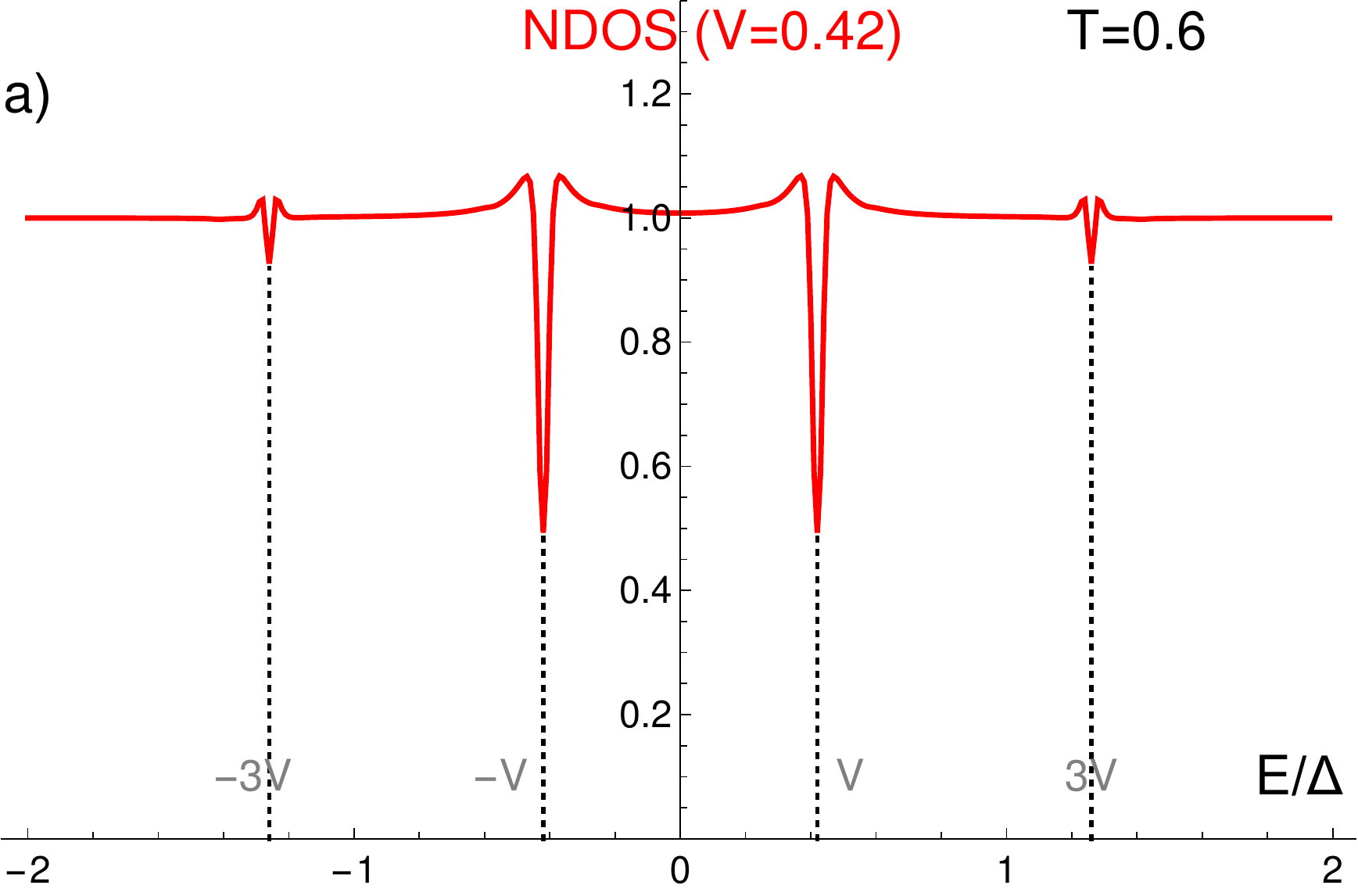} \hspace{0.2cm}
 \includegraphics[width=14pc]{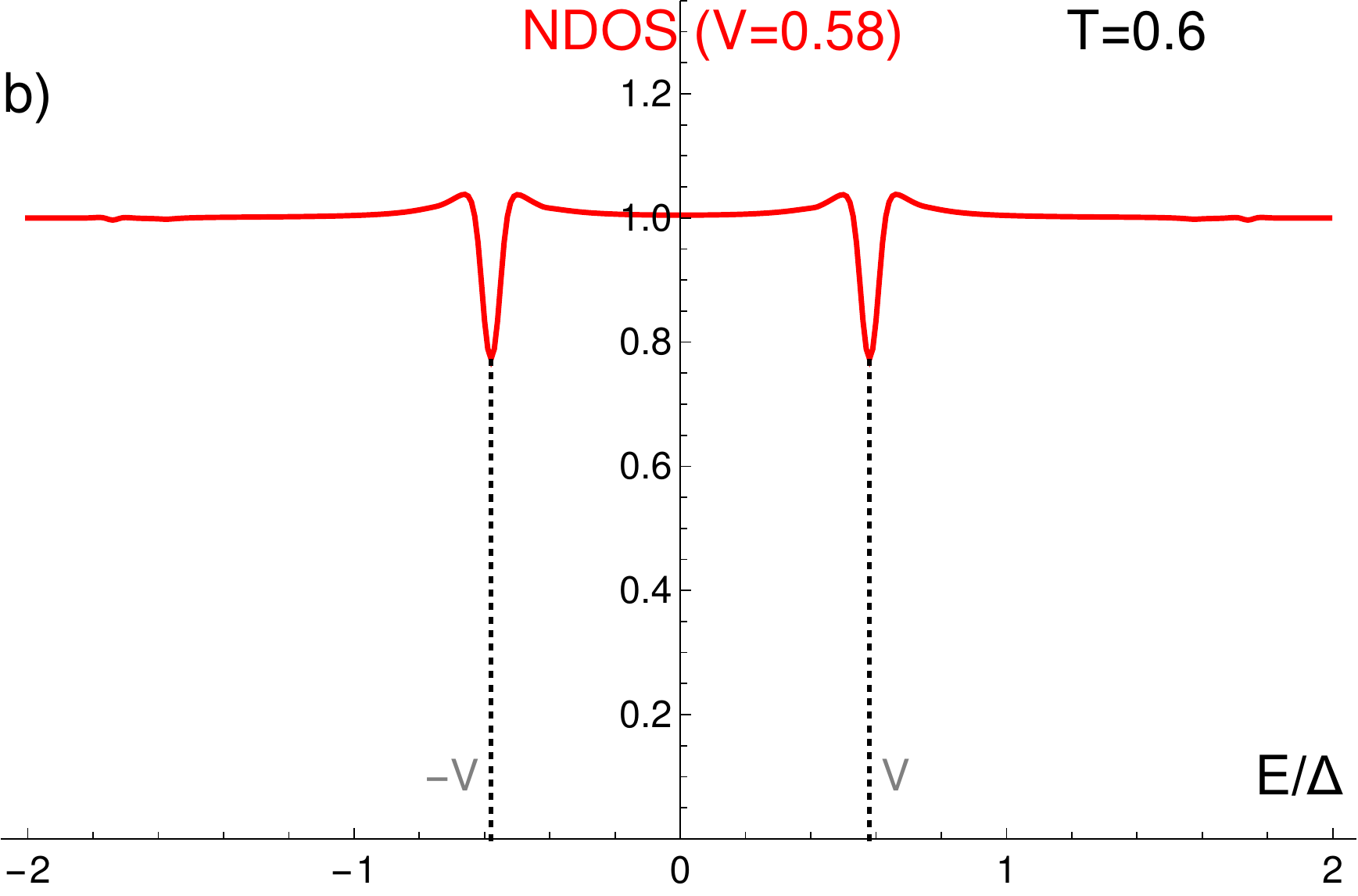}
 \caption{(Color online.) Same as Figs.~\ref{fig:DOSbiasedJJ1largetd}(a) and \ref{fig:DOSbiasedJJ1largetd}(b), but with transparency $T=0.6$.}
 \label{DOSbiasedJJ1_TR06}
\end{center} 
\end{figure}

When the two-terminal junction is biased, the NDOS shows a series of sharp resonances that are strongly affected by decoherence, as can be seen in Fig.~\ref{fig:DOSbiasedJJ1smalltd} ($\tau_d=0.05$) and in Fig.~\ref{fig:DOSbiasedJJ1largetd} ($\tau_d=5.0$) for small transparency, $T=0.1$.

For weak decoherence, Figs.~\ref{fig:DOSbiasedJJ1smalltd}(a) and \ref{fig:DOSbiasedJJ1smalltd}(b) show the peaks of the NDOS for voltages $V=0.42$ and $V=0.58$. The sharp peaks resemble the divergence of
the density of states of the bulk BCS-superconductor at the edges of the gap. 
The position of the peaks corresponds to the voltages, $-V$ and $V$. 
The transport in this regime is well described by coherent MAR. The peaks can be explained 
by a MAR diagram as in Fig.~\ref{figMARdiag}(a). 
The bulk gap edges of the two superconductors induce by proximity the peak structure of the junction NDOS. 
For instance, a quasiparticle leaving $S_1$ at energy $-\Delta+V$, 
is reflected as a hole in $S_3$ at energy $\Delta-3V$. Higher $p$-order MAR processes give rise to peaks 
at $\pm(\Delta\pm(2p+1)V)$. The dotted lines
in Figs.~\ref{fig:DOSbiasedJJ1smalltd}(a) and \ref{fig:DOSbiasedJJ1smalltd}(b) correspond to the position of the peaks predicted by the MAR diagram. 
They agree well with the computed NDOS. 
Larger structures appear, centered at $\pm (\Delta \pm V)$, weaker ones at $\pm(\Delta-3V)$, and 
increasingly weaker ones at $\pm(\Delta-5V)$, corresponding to the reduction of the MAR amplitude for transparency $T<1$. 
The linear dependence of the peaks position as a function of the bias voltage
$V$ is shown in Fig.~\ref{fig:DOSbiasedJJ1smalltd}(c). 
Four lines in the $E$-$V$ plane correspond to peaks of the NDOS at $\pm(\Delta \pm V)$, 
while weaker lines correspond to the higher order MAR at $\pm(\Delta-3V)$ and $\pm(\Delta-5V)$.

For strong decoherence, Fig.~\ref{fig:DOSbiasedJJ1largetd} shows a different structure of the NDOS. The peaks due to MAR processes are washed away and minigaps appear around energy $\pm V$, similar to the equilibrium minigap in Fig.~\ref{DOSeqJJ}(d). A similar minigap appears in diffusive SN junctions, with size given by the Thouless energy, where it is attributed to reflectionless tunneling \cite{reflectionless_tunneling, proximity_expt}.
The inset of Fig.~\ref{fig:DOSbiasedJJ1largetd}(a) shows the NDOS of an SNN interface, calculated using circuit theory, with voltage $V=0.45$ applied to the S electrode. The result shows the proximity minigap developing at the chemical potential of the S electrode, $E=V$. Due to strong decoherence, the NDOS of the SNS junction exhibits the separate signature of each NS interface.

\begin{figure*}[t!]
 \centerline{\includegraphics[width=14pc]{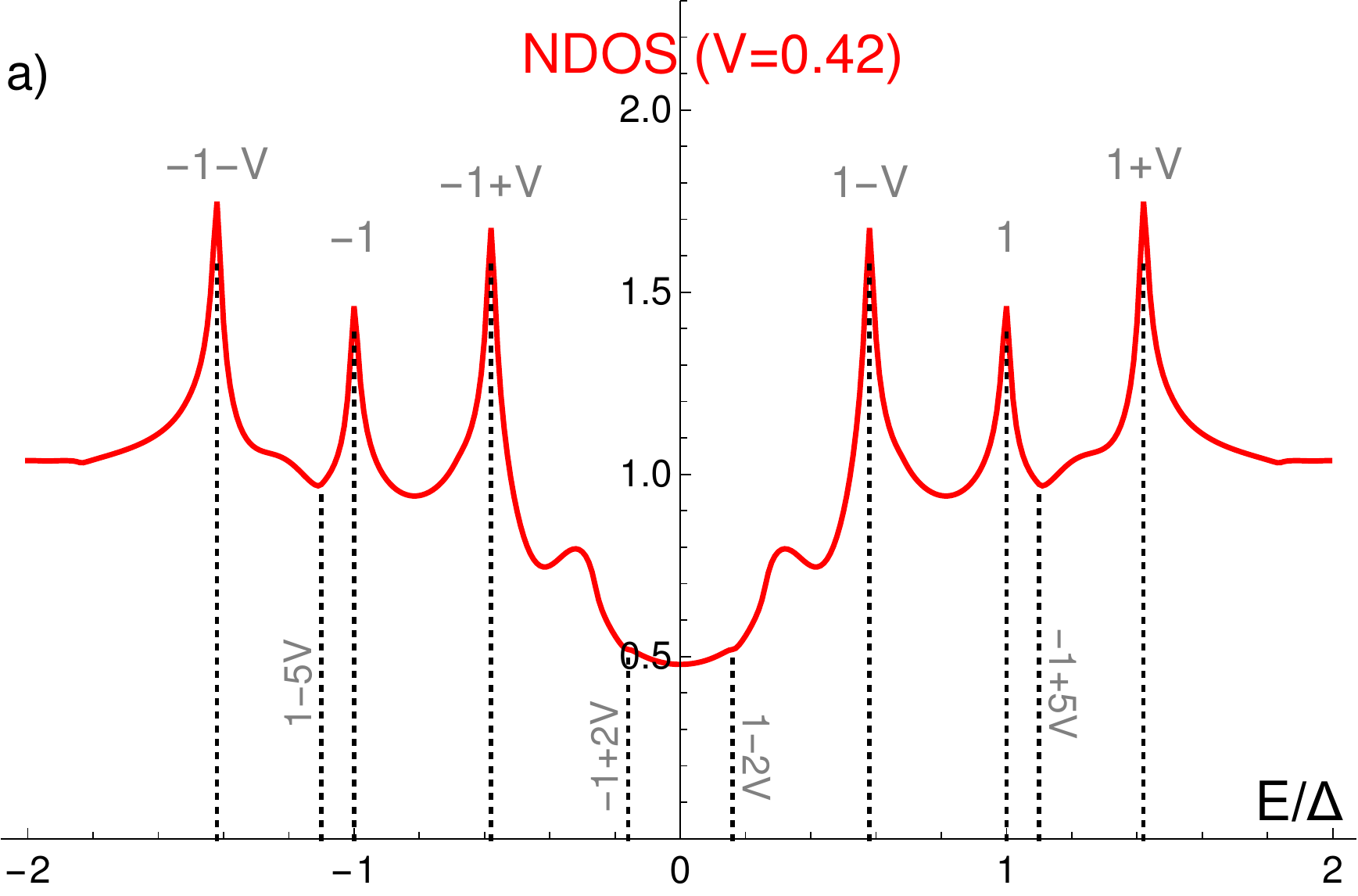} \hspace{0.1cm}
 \includegraphics[width=14pc]{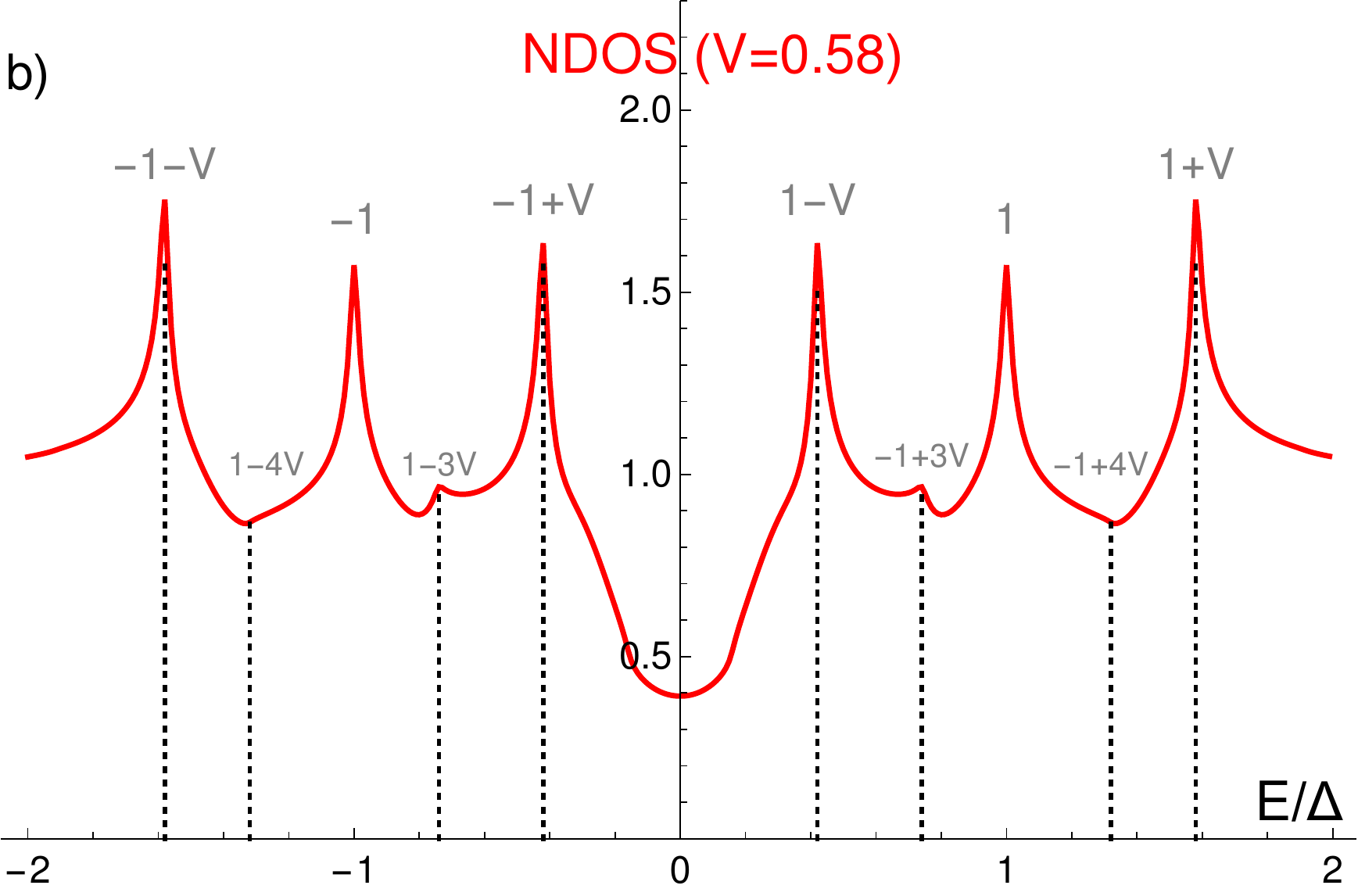}\hspace{0.2cm}
 \includegraphics[width=10pc]{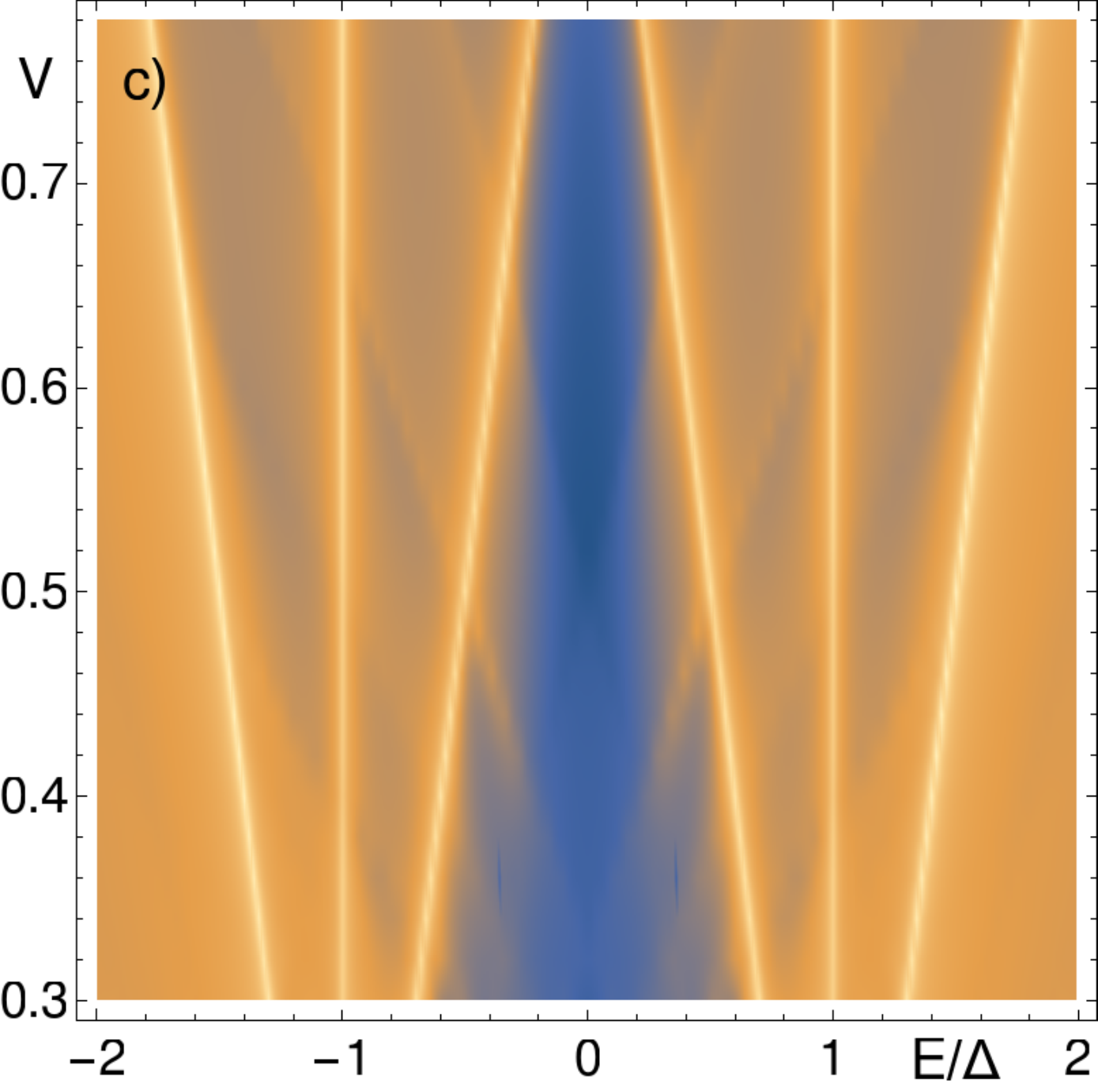}}
 \caption{(Color online.) Normal density of states (NDOS) as a function of energy $E$ in a biased TTJ with three electrodes at voltages $-V,0,V$, at transparency $T=0.1$,
  for small decoherence ($\tau_d=0.05$). (a) and (b) show the NDOS for $V=0.42$ and 
  $V=0.52$, and (c) shows the density plot of the NDOS as a function of the energy $E$ and the voltage $V$. In (a) and (b), the vertical dotted lines show the positions of the expected peaks at voltages $\pm(\Delta\pm(2p+1)V)$ or $\pm(\Delta\pm(2p)V)$ due to MAR processes. ($\Delta=\hbar=e=1$.)}
 \label{fig:DOSbiased3S1smalltd}
\end{figure*}

\begin{figure*}[t!]
 \centerline{\includegraphics[width=14pc]{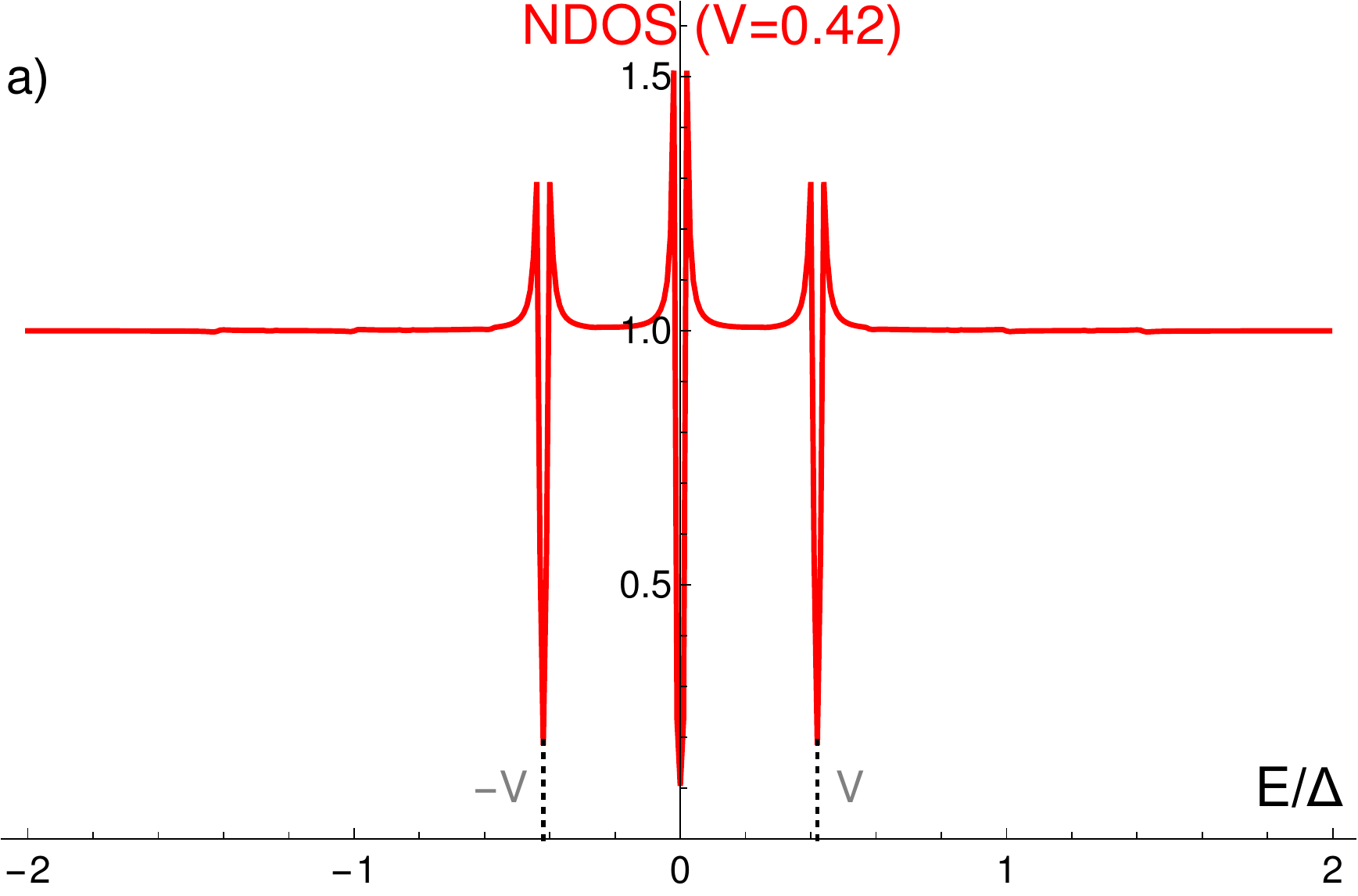} \hspace{0.1cm}
 \includegraphics[width=14pc]{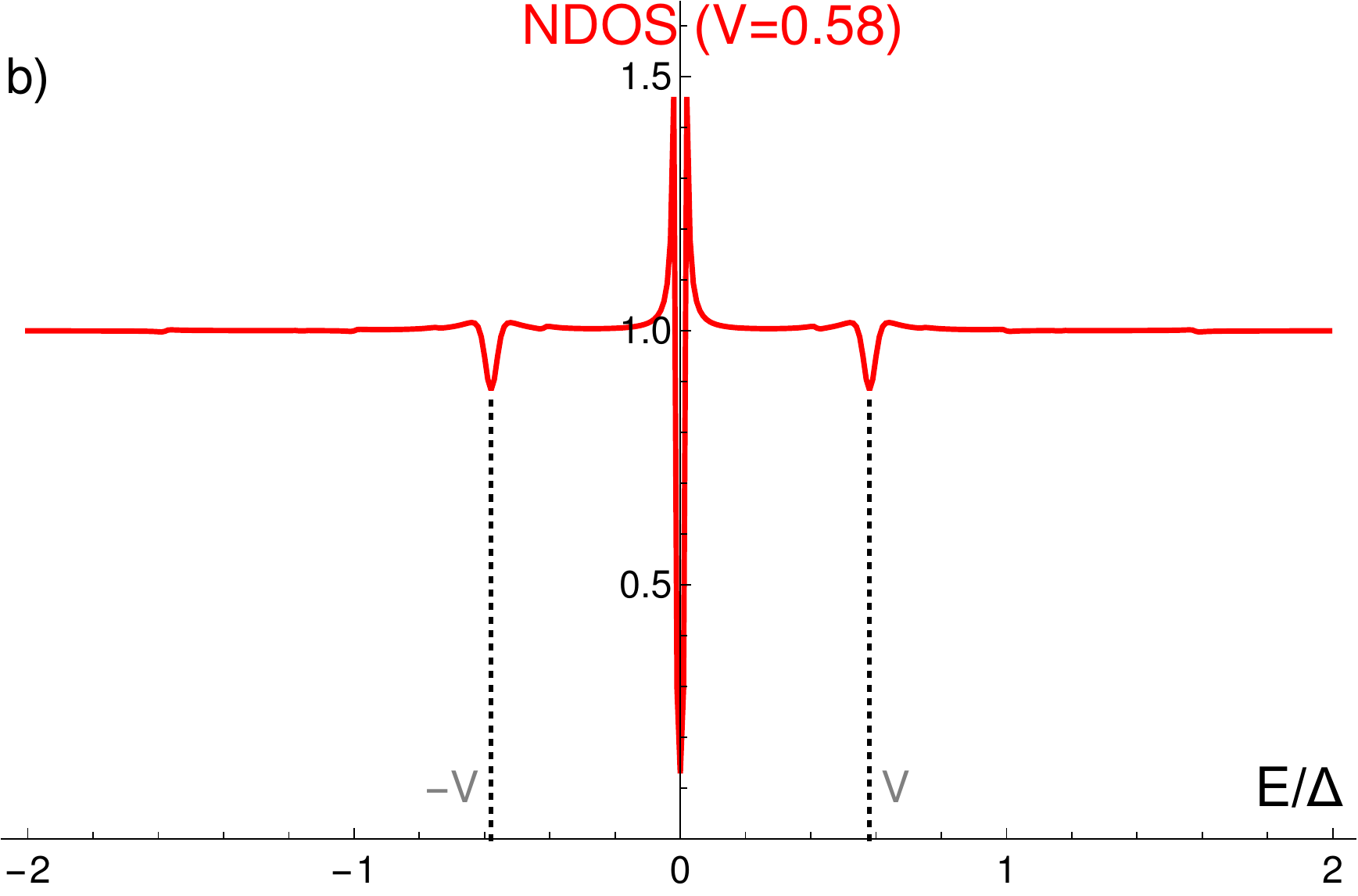}\hspace{0.2cm}
 \includegraphics[width=10pc]{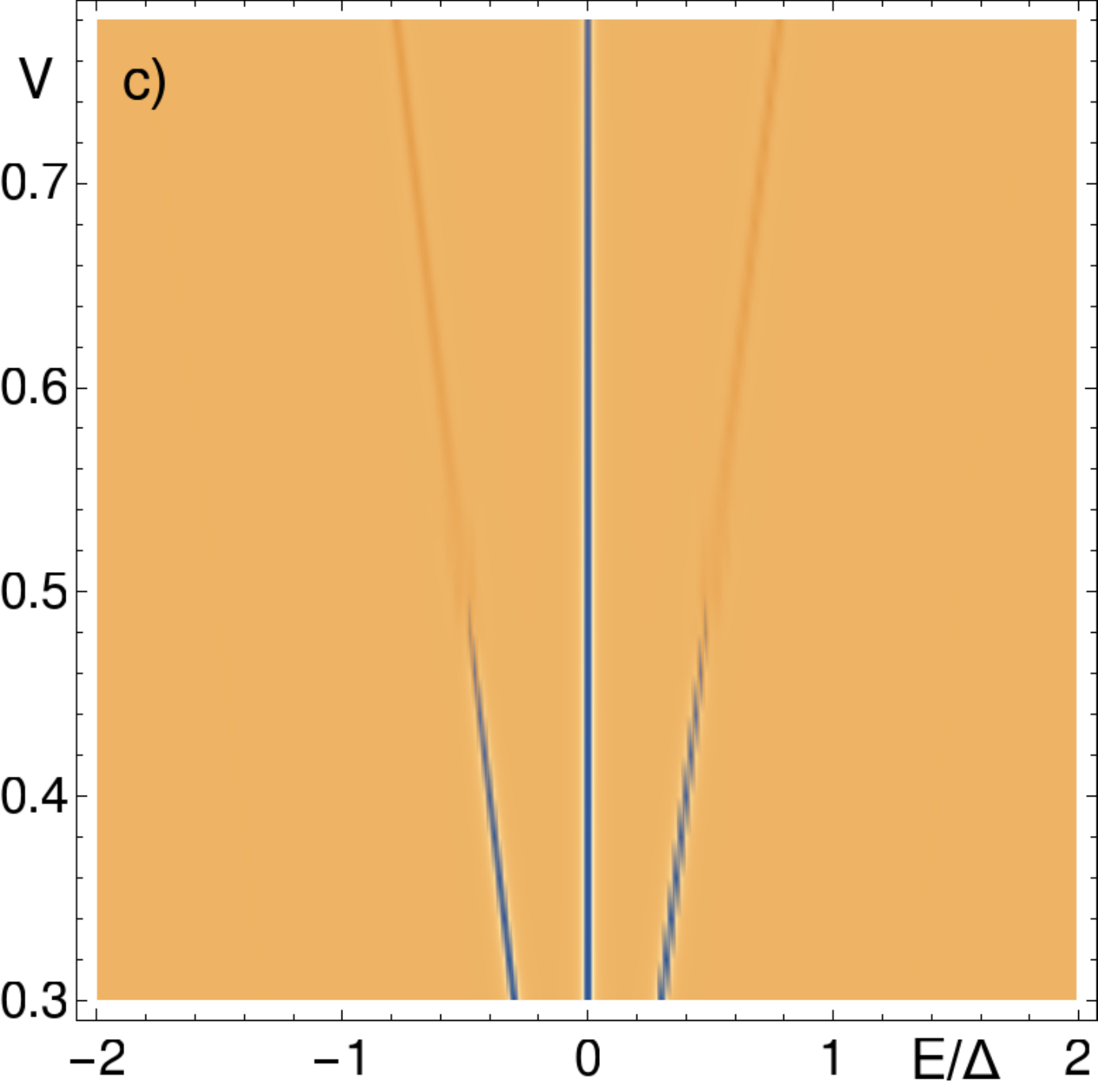}}
 \caption{(Color online.) Same as Fig.\ref{fig:DOSbiased3S1smalltd}, but for large decoherence ($\tau_d=5.0$).
 In (a) and (b), the vertical dotted lines show the positions of the expected minigaps peaks at voltages $\pm V$ and 0 corresponding to the chemical potentials of the two electrodes.}
 \label{fig:DOSbiased3S1largetd}
\end{figure*}

Two qualitatively different regimes are observed in Fig.~\ref{fig:DOSbiasedJJ1largetd}(c) for $V<\Delta/2$ and $V>\Delta/2$. At the threshold, the lower edge of the bulk gap of the terminal biased at $+V=\Delta/2$ coincides with the chemical potential of the terminal biased at $-V=-\Delta/2$. The same condition marks the threshold of the $n=2$ MAR process, as seen in Fig.~\ref{figMARdiag}(b). The regimes differ in the coupling strength to the continuum of quasiparticle states, the regime $V>0.5 \Delta$ marking the stronger coupling giving rise to rounded features of the NDOS.  

In Fig.~\ref{DOSbiasedJJ1_TR06}(a), we observe small corrections to the structure of the NDOS, visible at large transparency, $T=0.6$. The feature can be ascribed to the $n=4$ MAR process and corresponds to energy $E=\pm 3V$. In contrast, the transparency does not significantly modify the NDOS in the case of small decoherence (for this reason we only show the results for $T=0.1$). Indeed, for a symmetric two-terminal junction and for vanishing $\tau_d$, the small transparency limit of the circuit theory equations has been previously shown to be identical to the exact result \cite{VanevicBelzig2006}.



\begin{figure*}[t]
 \centerline{\includegraphics[height=10pc]{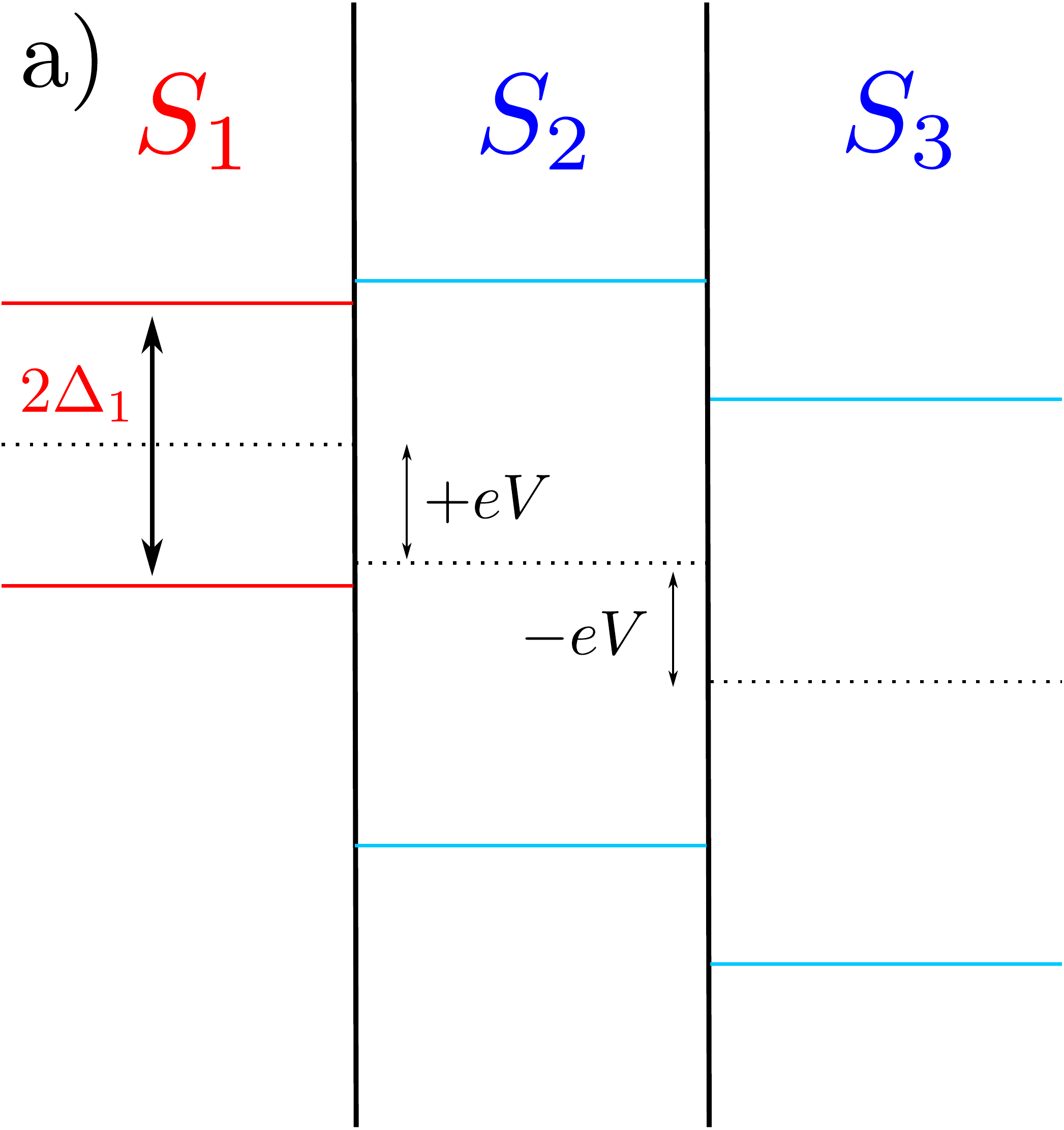} \hspace{1.0cm}
 \includegraphics[height=10pc]{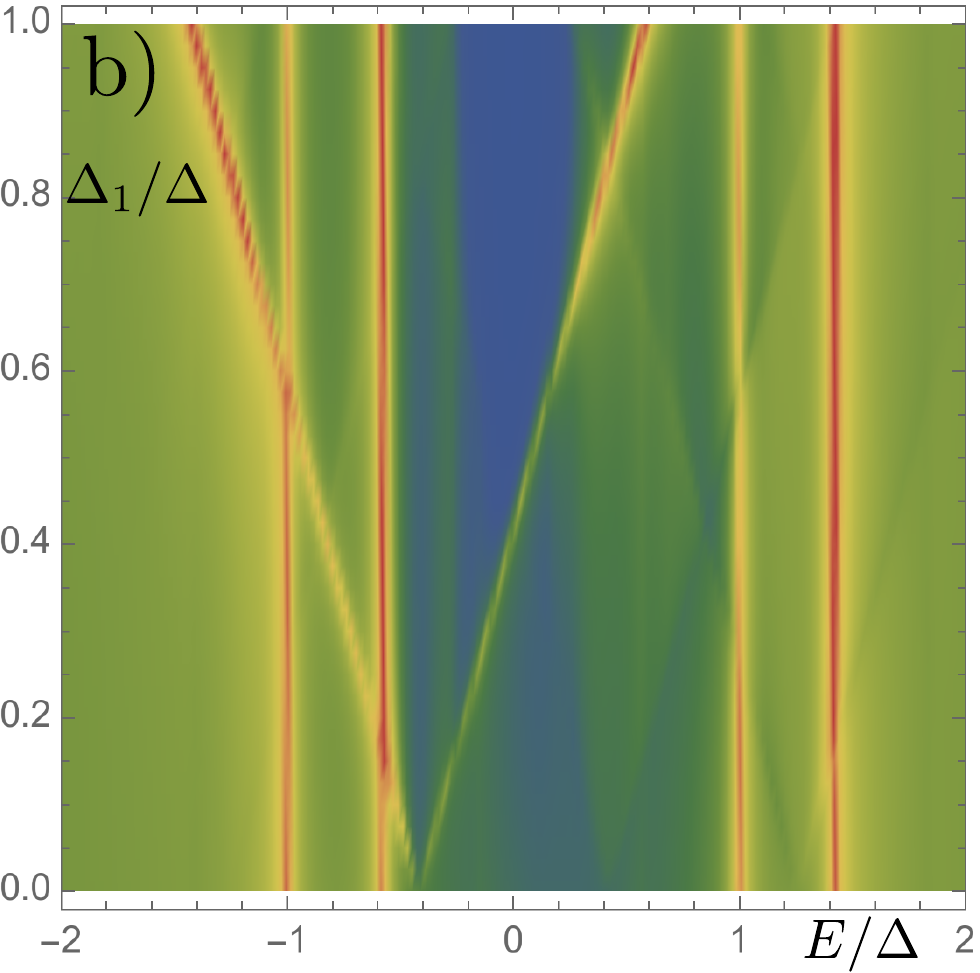} \hspace{1.0cm}
 \includegraphics[height=10pc]{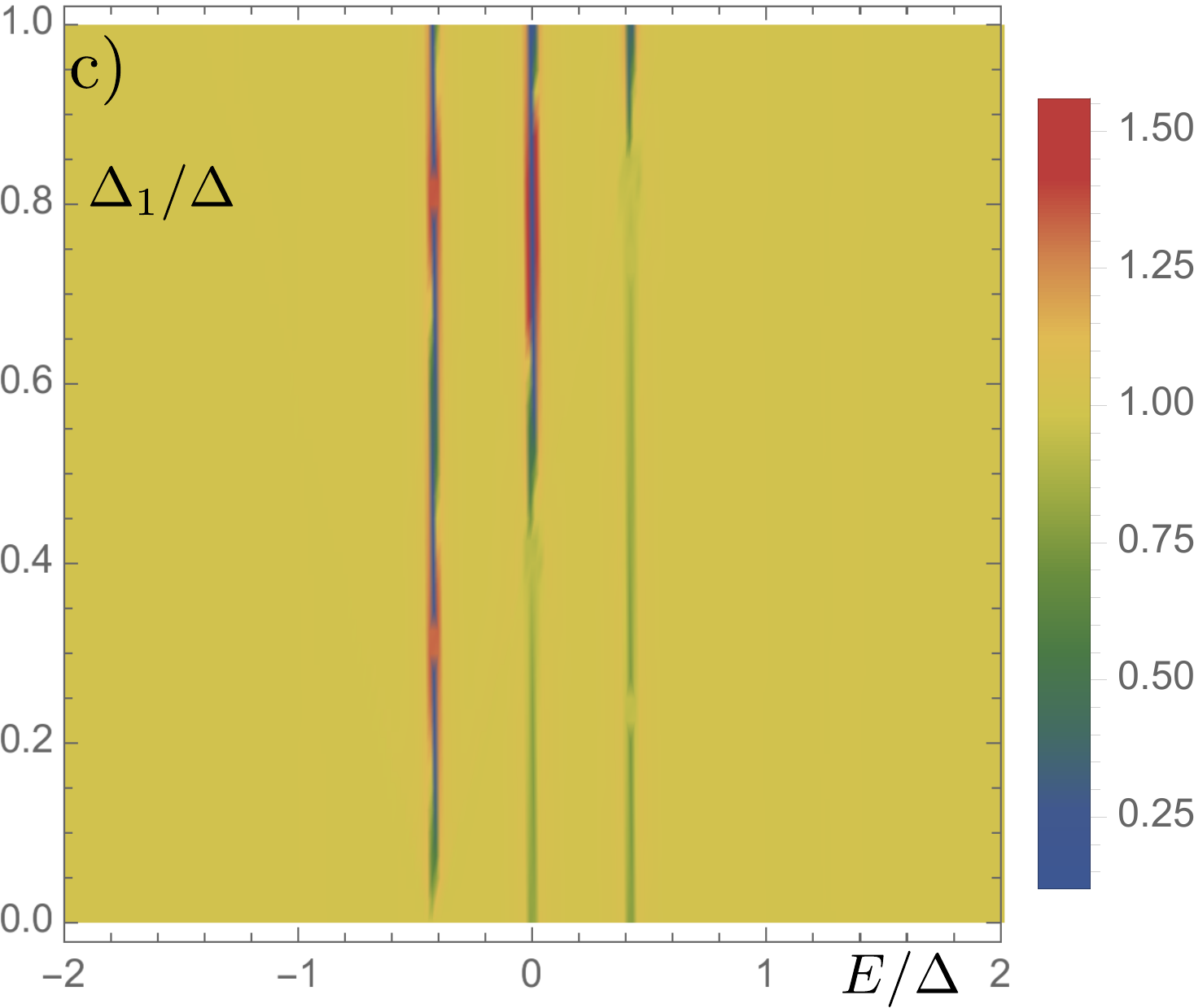}}
 \caption{(Color online.) NDOS as a function of energy $E/\Delta$ 
for asymmetric bulk superconducting gaps, at transparency $T=0.1$ and voltage $V=0.42$, 
(a) schematic illustration of asymmetric gap $\Delta_1$, with $\Delta_2=\Delta_3=\Delta$, (b) at $\tau_d=0.05$ as a function of $\Delta_1$, 
(c) at $\tau_d=5$ as a function of $\Delta_1$. ($\Delta=\hbar=e=1$.)}
 \label{fig:NDOSvsDelta1}
\end{figure*}

\begin{figure*}[t]
 \centerline{\includegraphics[height=10pc]{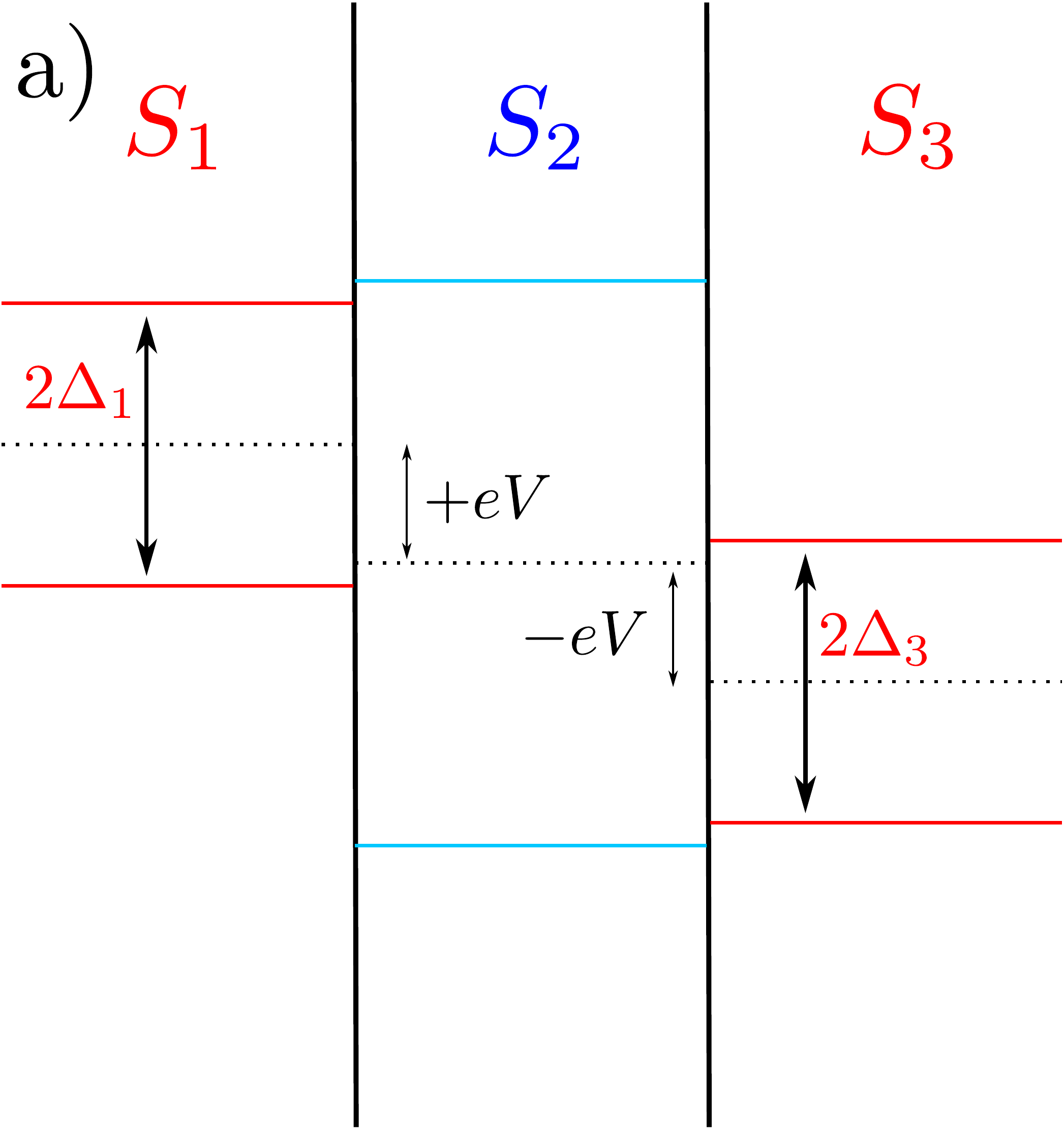} \hspace{1.0cm}
 \includegraphics[height=10pc]{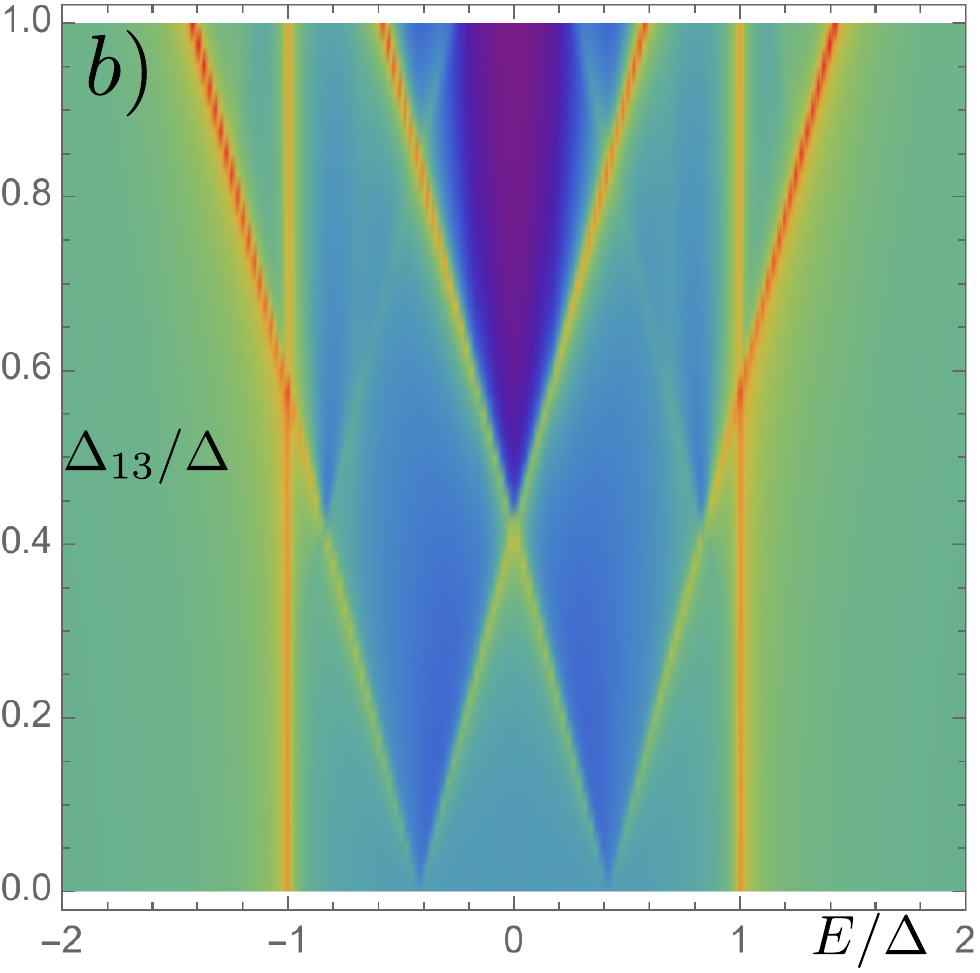} \hspace{1.0cm}
 \includegraphics[height=10pc]{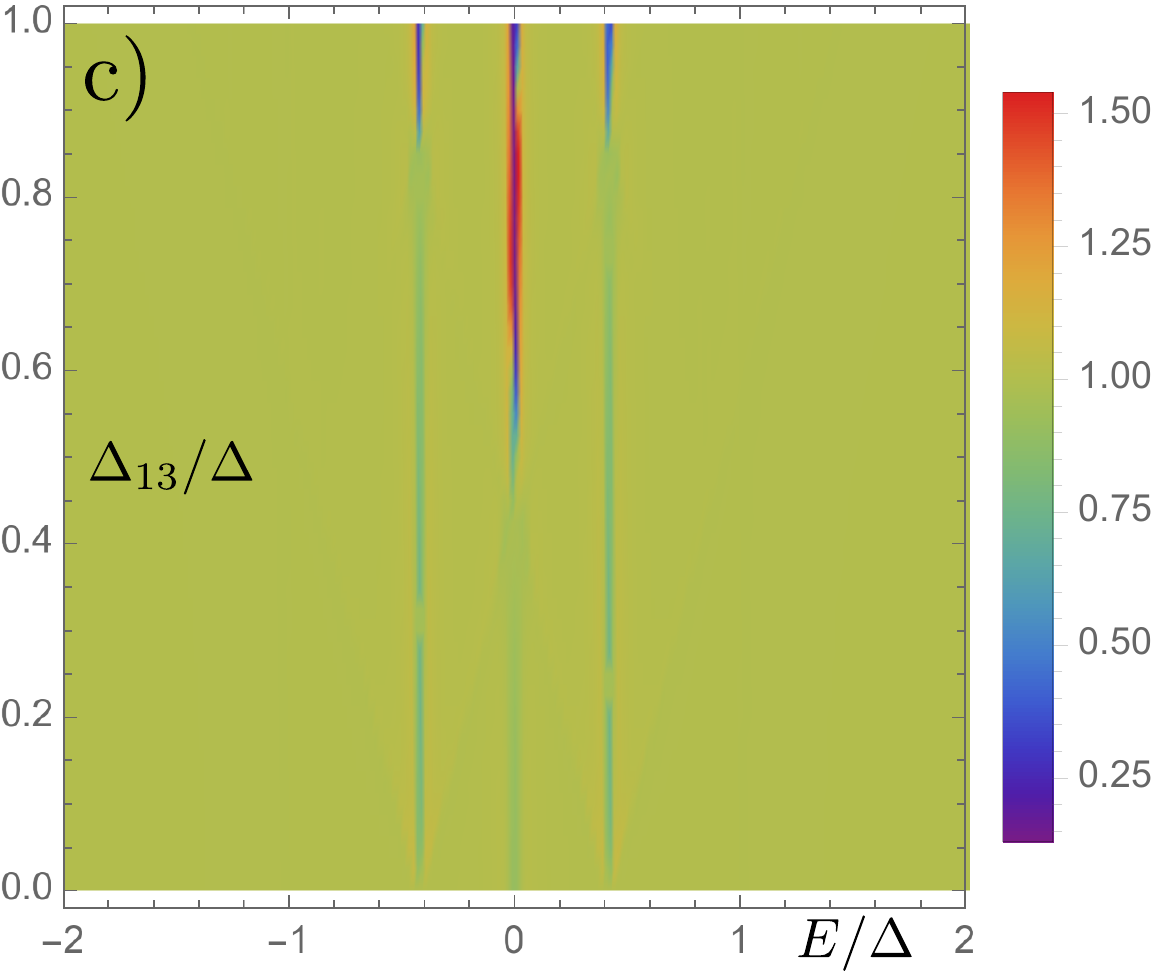}
 }
 \caption{(Color online.) Same as Fig. \ref{fig:NDOSvsDelta1}, 
(a) schematic illustration of asymmetric gaps $\Delta_1=\Delta_3=\Delta_{13}$, with $\Delta_2=\Delta$, (b) at $\tau_d=0.05$ as a function of $\Delta_{13}$, and (c) at $\tau_d=5$ as a function of $\Delta_{13}$.}
 \label{fig:NDOSvsDelta13}
\end{figure*}

\begin{figure}[t]
\begin{center}
\includegraphics[width=\linewidth]{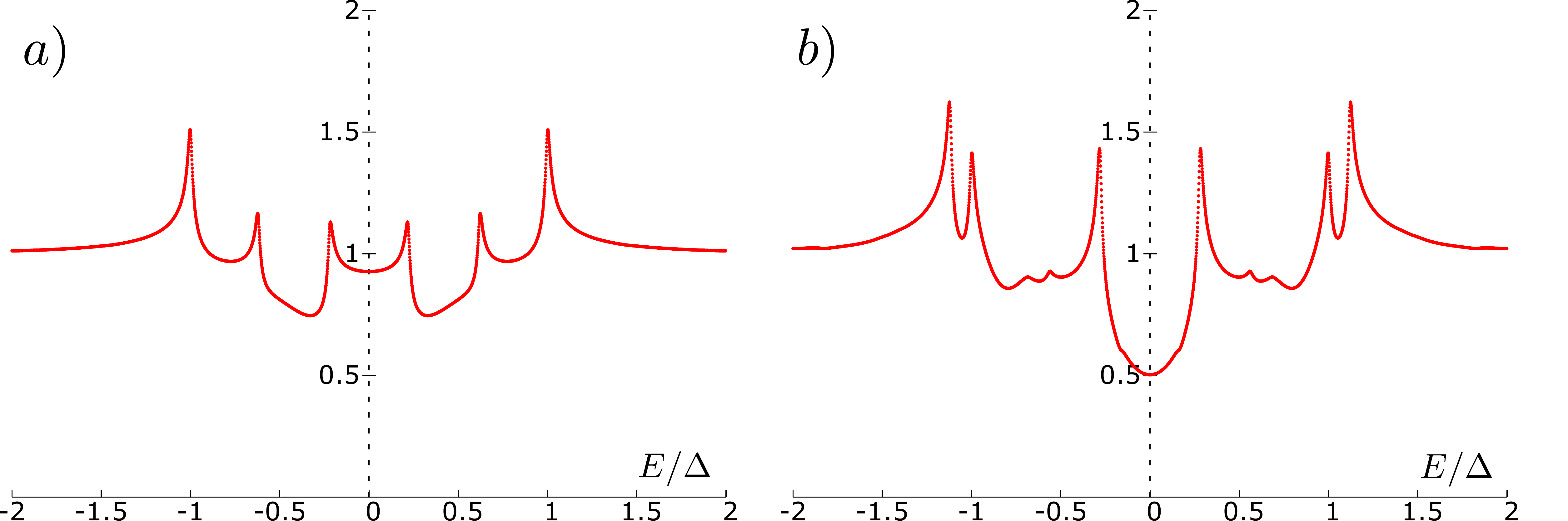}
\caption{Line cut of Fig.~\ref{fig:NDOSvsDelta13}(b) at fixed $\Delta_{13}$ for (a) $\Delta_{13}=0.2 \Delta$ and (b) $\Delta_{13}=0.7 \Delta$.}
\label{linecuts}
\end{center}
\end{figure}

\section{The three-terminal junction}
\label{biasedTTJ}

We now consider a metallic TTJ with superconducting terminals $S_{i}$, $i=\{1,2,3\}$, 
biased at voltages $V_1=-V_3=V$, $V_2=0$. 
Below, we set all superconducting phases to zero and discuss the effect of bias, transparency
and bulk gap asymmetry on the structure of the NDOS. In the following subsection, we discuss the dependence
of the NDOS on the phase $\varphi_Q$. 

\subsection{NDOS at $\varphi_Q=0$}

For a symmetric junction, $\Delta_i=\Delta$ and $T_i=T$, we find that the junction transparency
does not significantly modify the spectral structure of the NDOS, with few exceptions that we will point out. 
For this reason, we choose $T=0.1$ to produce the plots.
The NDOS of the TTJ in the limits of small and large decoherence, Figs.~\ref{fig:DOSbiased3S1smalltd}
and \ref{fig:DOSbiased3S1largetd}, respectively, are similar to those for the conventional,
two-terminal junction, presented in Figs.~\ref{fig:DOSbiasedJJ1smalltd} and  
\ref{fig:DOSbiasedJJ1largetd}, with additional structures emerging from the presence of a third
terminal, biased at $V_2=0$.

\begin{figure*}[t!]
 \centerline{
 \includegraphics[height=10pc]{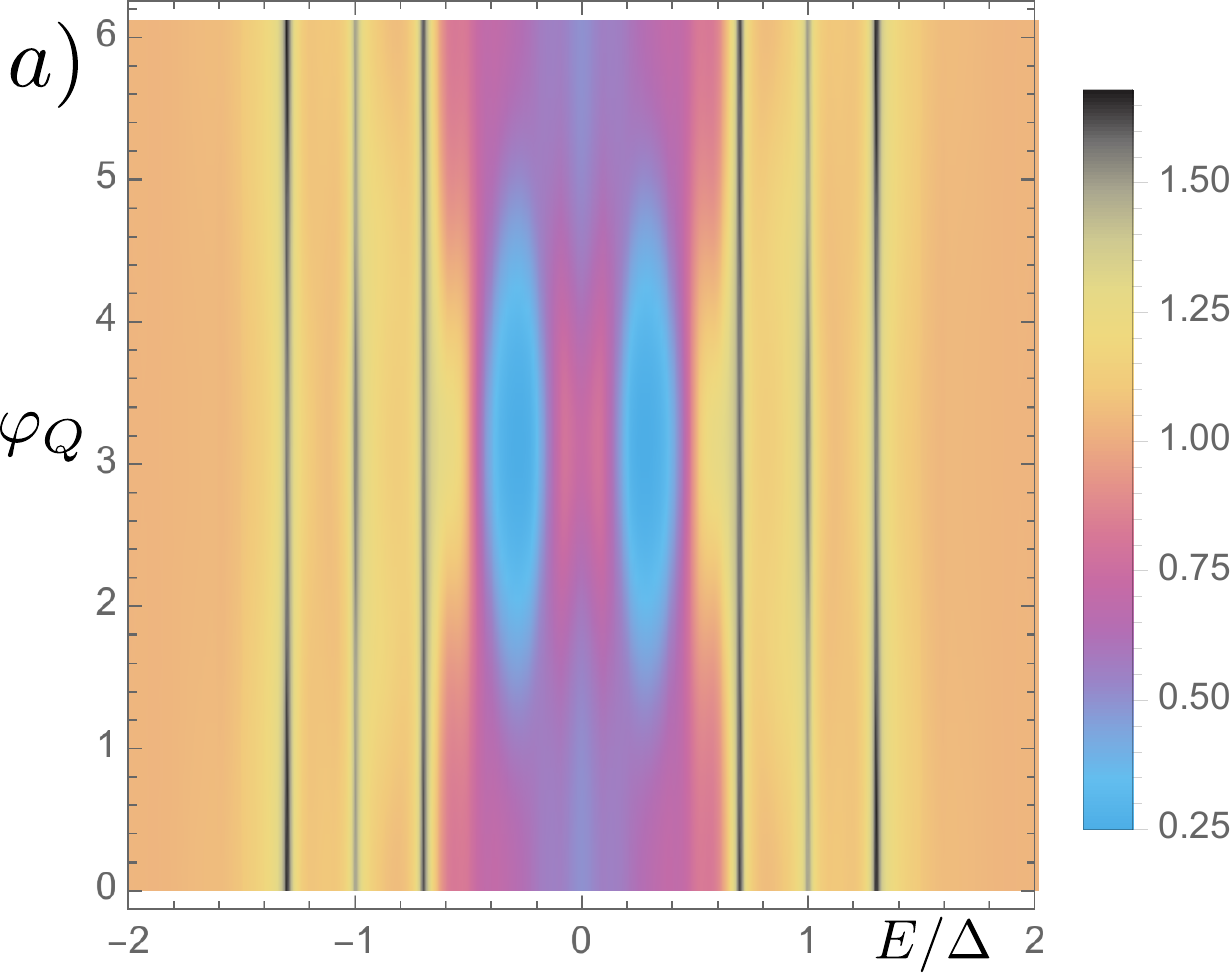} \hspace{0.3cm}
 \includegraphics[height=10pc]{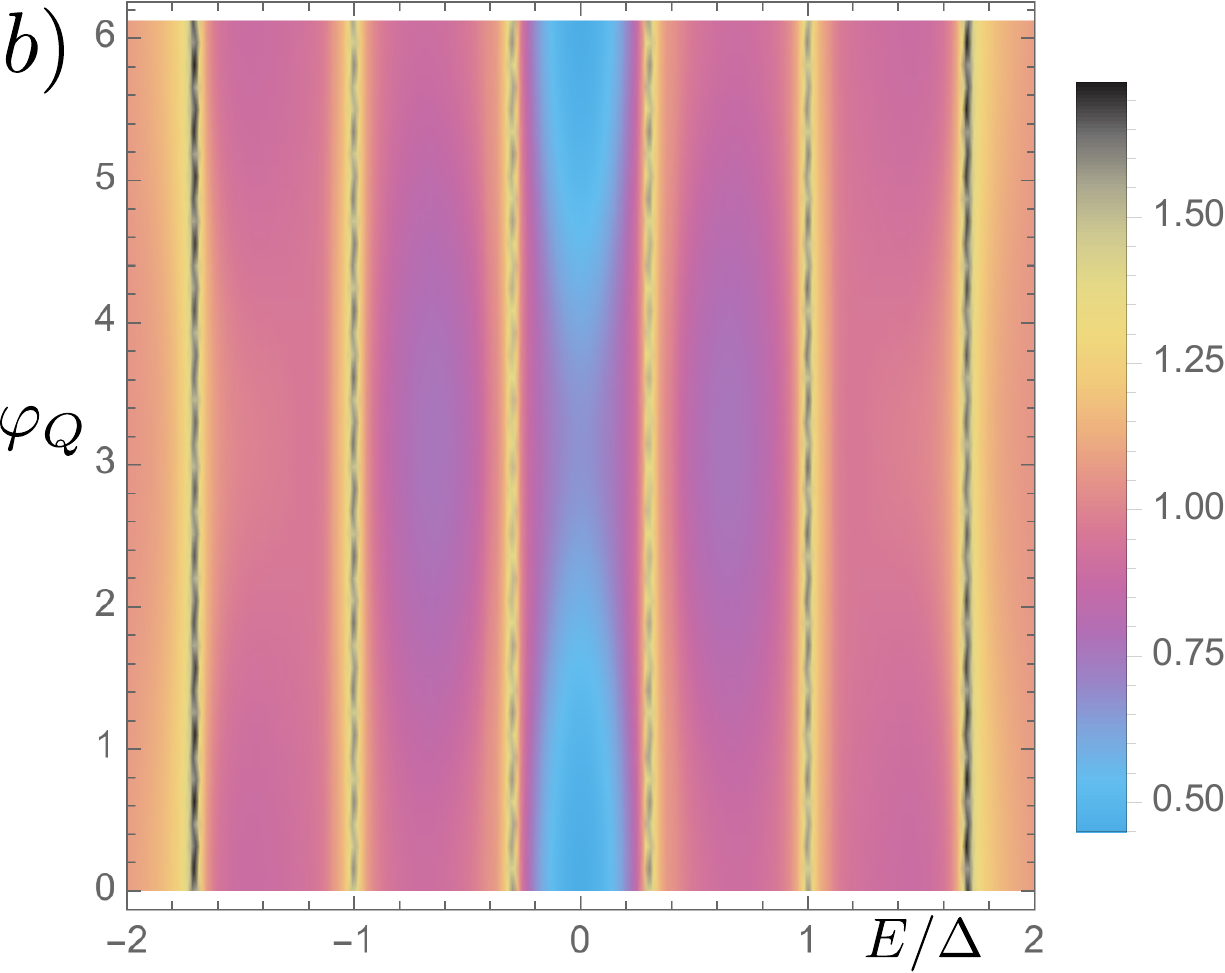} \hspace{0.3cm}
 \includegraphics[height=10pc]{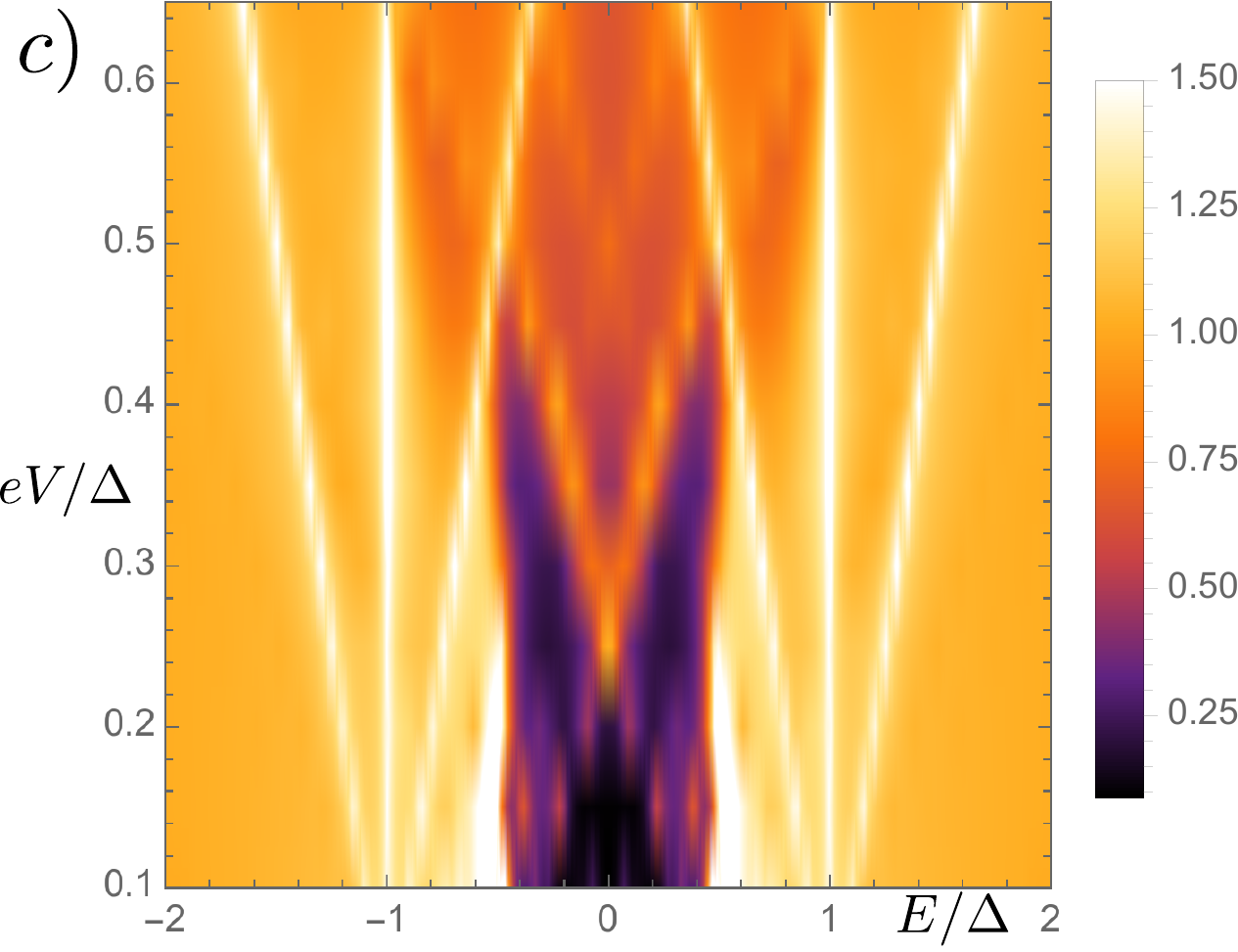}}
 \caption{(Color online.) NDOS as a function of energy $E/\Delta$ 
in a biased TTJ with three electrodes at voltages $-V,0,V$,
for small decoherence  $\tau_d=0.05$ and transparency $T=0.1$, 
(a) at $V=0.3$ as a function of $\varphi_Q$, 
(b) at $V=0.7$ as a function of $\varphi_Q$, and
(c) at $\varphi_Q=\pi$ as a function of $V$. ($\Delta=\hbar=e=1$.)}
 \label{fig:NDOS_PhiQ=Pi}
\end{figure*}

\begin{figure*}[t!]
\begin{center}
\begin{tabular}{cc}
\includegraphics[width=16pc]{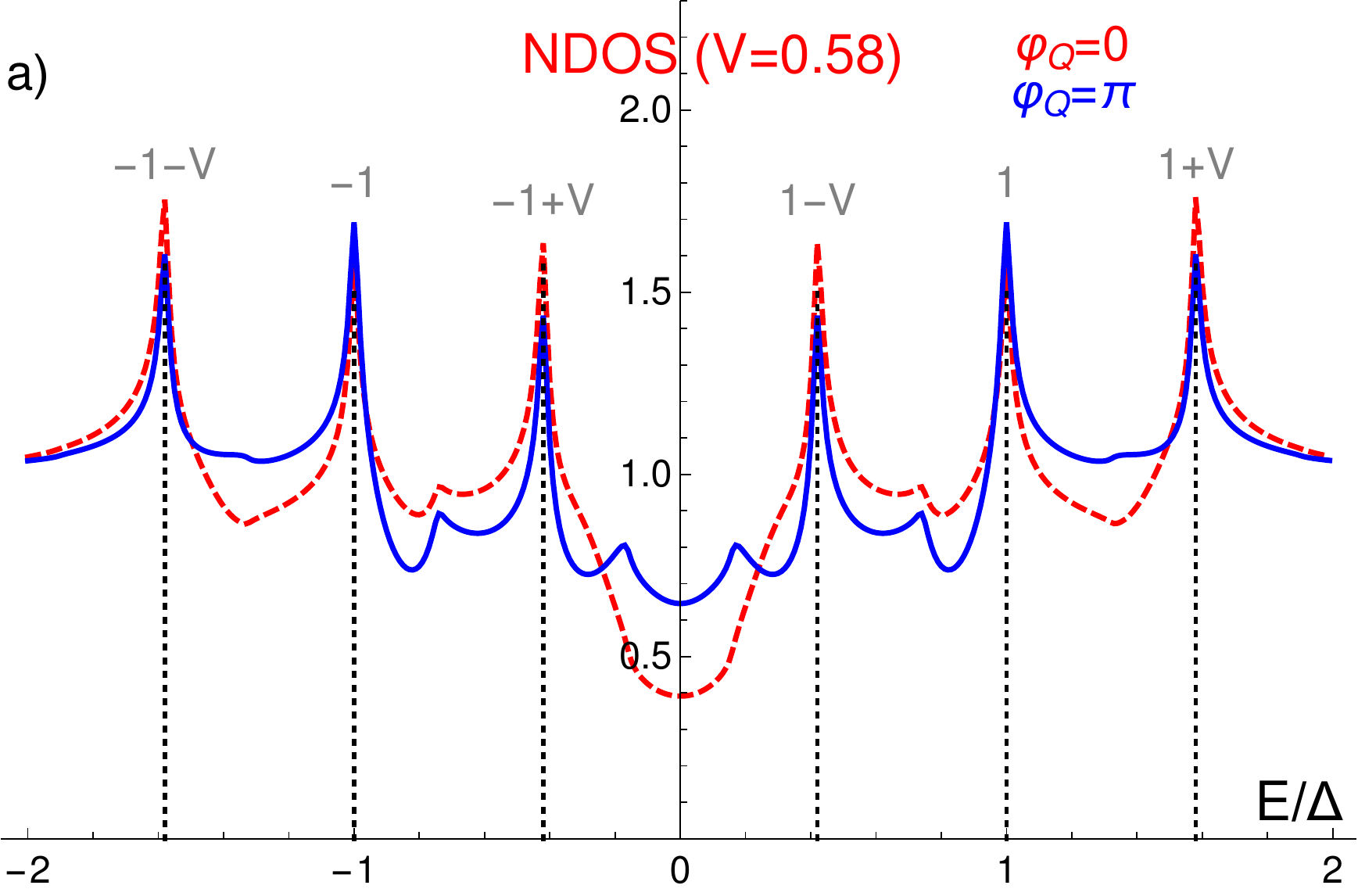} &
\hspace{1.cm} \includegraphics[width=16pc]{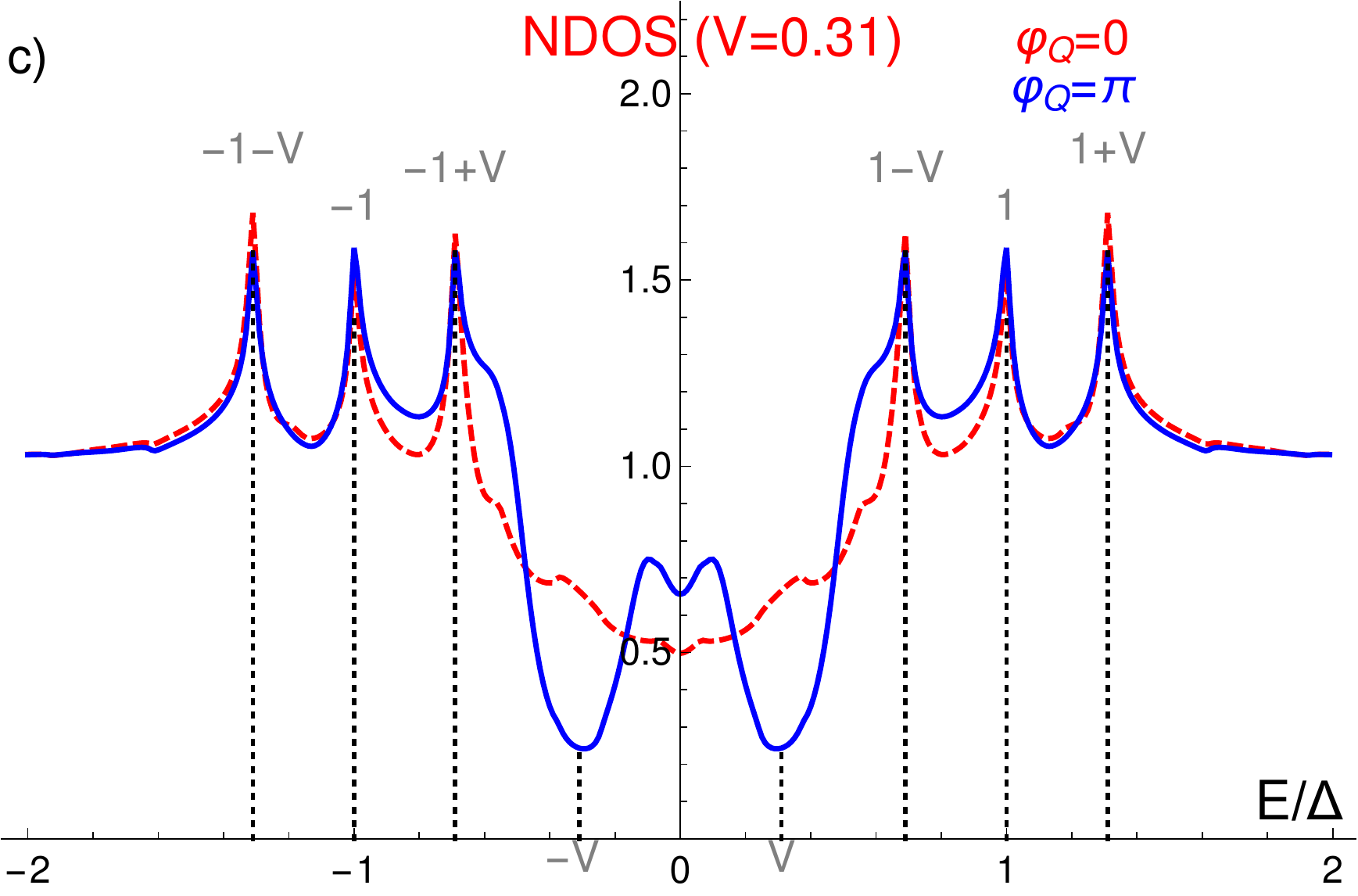} \\
\includegraphics[width=16pc]{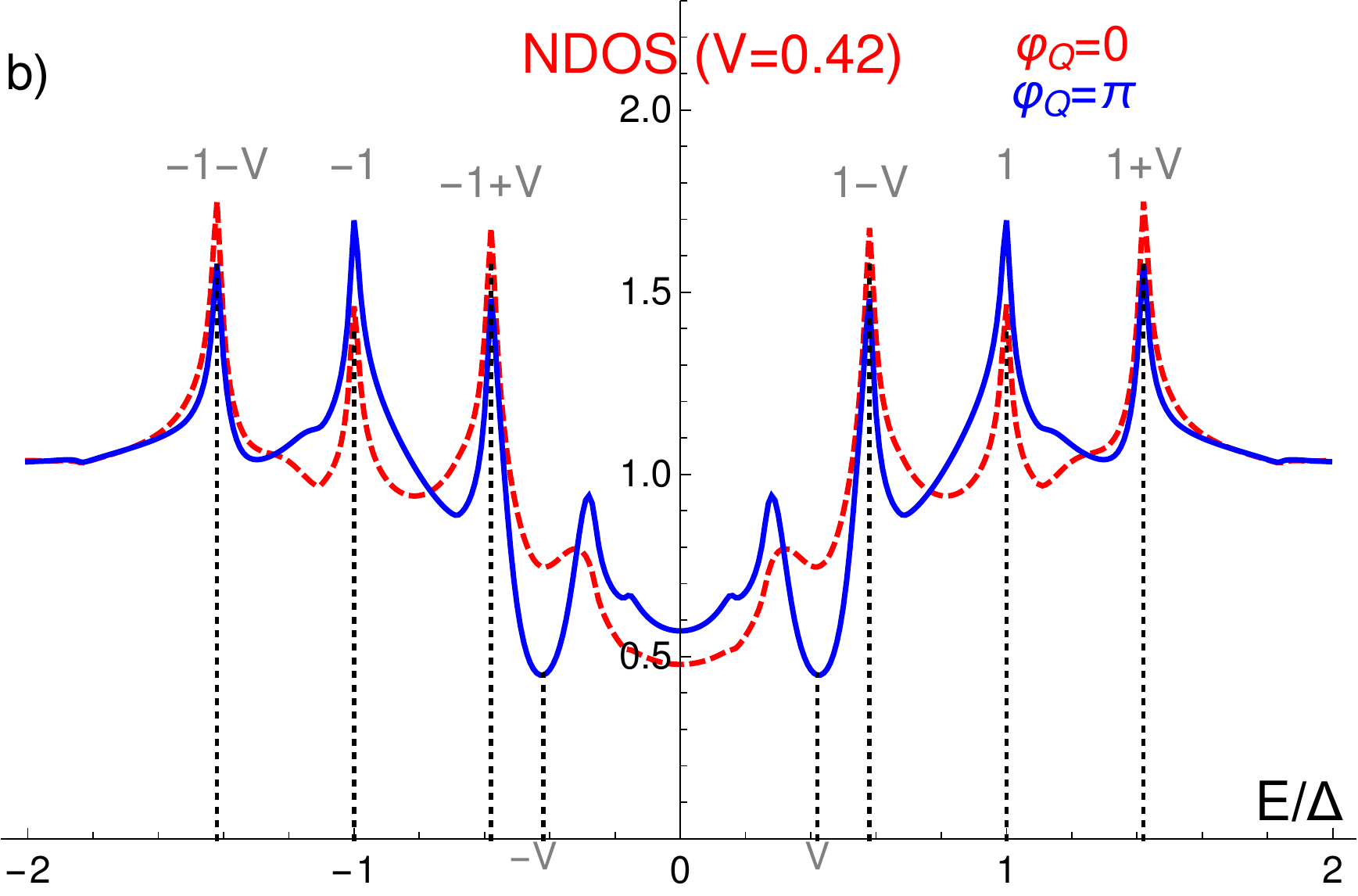} & 
\hspace{1.cm}\includegraphics[width=16pc]{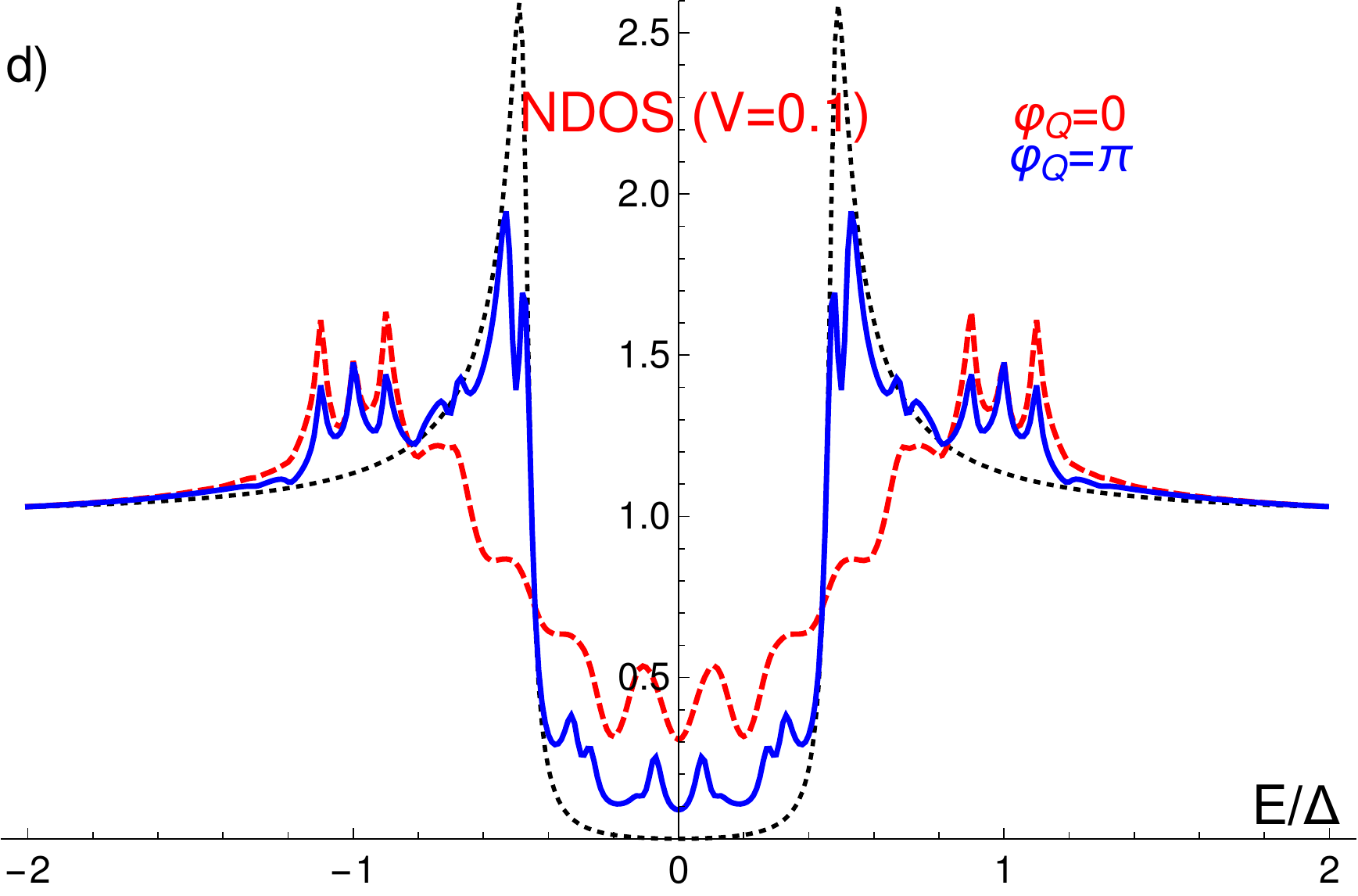}
\end{tabular}
\end{center}
 \caption{(Color online.) NDOS as a function of energy $E$ 
in a biased TTJ with three electrodes at voltages $-V,0,V$,
 for small decoherence  $\tau_d=0.05$ and transparency $T=0.1$, for (a)
  $V=0.58$, (b) $V=0.42$, (c) $V=0.31$, and (d) $V=0.1$.
   The red (dashed) curves correspond
 to $\varphi_Q=0$, while the blue (full) curves
 correspond to $\varphi_Q=\pi$. ($\Delta=\hbar=e=1$.)}
 \label{fig:DOSbiased3SPhiPi}
\end{figure*}

For small
decoherence, Fig.~\ref{fig:DOSbiased3S1smalltd},  
the NDOS presents an additional structure originating from the gap edges
of the superconducting electrode at voltage $V_2=0$ at $\pm (\Delta+V_2)$. 
New MAR channels develop: quasiparticles with energy near the gap edges in $S_{1,3}$ scatter at 
energies $\pm  (\Delta\pm(2p+1)V)$, while those with energy near 
the gap edge in $S_{2}$ scatter at energies $\pm (\Delta\pm 2pV)$. 
Corresponding structures of the NDOS can be seen in Fig.~\ref{fig:DOSbiased3S1smalltd}
for all MAR processes of order $n$ smaller than $n<(V/\Delta)$. The NDOS presents six large peaks, 
at energies $\pm(\Delta \pm V)$ and $\pm \Delta$ that correspond to the gap edges, 
and smaller structures corresponding to higher-order MAR processes. 
The position of these peaks depends linearly on the voltage as seen in  Fig.~\ref{fig:DOSbiased3S1smalltd}(c), 
in agreement with the interpretation in terms of MAR diagrams.

For large decoherence, Fig.~\ref{fig:DOSbiased3S1largetd}, the NDOS presents an additional (third)
minigap at $E=V_2=0$, compared to the two-terminal junction. The third minigap is unaffected by the two
voltage regimes $V \lessgtr \Delta/2$ that determine the shape of the minigap edges at $E=\pm V$. 
The edges of the minigap at $E=0$ remain sharp irrespective of $V$, since the chemical potential 
of the corresponding superconductor, $S_2$, does not change with $V$.

It is interesting to track the position of the resonant structures of the NDOS by introducing an 
asymmetry in the bulk superconducting gaps. If $S_1$ is a weaker superconductor with 
$\Delta_1<\Delta$, Fig. \ref{fig:NDOSvsDelta1}(b) shows the linear dependence of the corresponding MAR resonance on $\Delta_1$,
confirming its relation to the gap edge of $S_1$, according to the expression $\pm (\Delta_1\pm2pV)$.
In the large decoherence regime, the position of minigaps does not change with $\Delta_1$. However, the shape
of the minigap corresponding to the $NS_1$ interface is strongly affected by $\Delta_1$, showing the transition
between sharp and smoothed out features at the threshold $V=\Delta_1/2$. The gap $\Delta_1$ has an effect on the 
shape of the other two minigaps as well, signaling the persistence of weak non-local transport in the large 
decoherence regime.

Similarly, if both biased superconductors, $S_1$ and $S_3$, have a smaller bulk gap, $\Delta_1=\Delta_3\equiv \Delta_{13}$, with
$\Delta_{13}<\Delta_2=\Delta$, the corresponding resonances shift linearly with $\Delta_{13}$, 
as seen in  Fig. \ref{fig:NDOSvsDelta13}.
For low decoherence, the asymmetric setup shares similarities with the well-known
Cooper pair splitter (CPS) setup \cite{lesovik2001,recher2001}, where a junction between a superconductor
and two normal metals is formed, the metals acting as collectors for 
electrons resulting from splitting Cooper pairs. 
In our setup, the collectors
are the weaker superconductors $S_1$ and $S_3$, and the CPS device is recovered
in the limit $\Delta_{13}\rightarrow 0$. From Fig.~\ref{fig:NDOSvsDelta13}
we observe that in the regime where $\Delta_{13}<V$, the modification
to the CPS NDOS mainly consists of small pseudogaps located at $\pm V$, of width
$2\Delta_{13}$. 
These are similar to the proximity-induced minigap at a biased SN interface, 
see the inset of Fig.~\ref{fig:DOSbiasedJJ1largetd}(a), and their positions match the
chemical potential of superconductors $S_1$ and $S_3$.

A qualitatively different regime is obtained in the opposite limit
$\Delta>\Delta_{13}>V$, where a pronounced pseudogap in the NDOS opens at zero energy,
with width $2(\Delta_{13}-V)$.
The gap corresponds to the energy window where coherent
transport processes of the quartet type participate in the transport. In this regime,
non-local processes of the quartet type coexist with Cooper pair splitting processes,
both mechanisms contributing to non-local correlations in the currents $I_1$ and $I_3$.
The comparison is shown in Fig.~\ref{linecuts} where
(a) shows a line cut at small $\Delta_{13}$, while (b) shows the
gap around $E=0$ that develops for large $\Delta_{13}$.
The complex statistical properties of transport in this regime will be discussed elsewhere.

\begin{figure}[t!]
\begin{center}
\includegraphics[width=15pc]{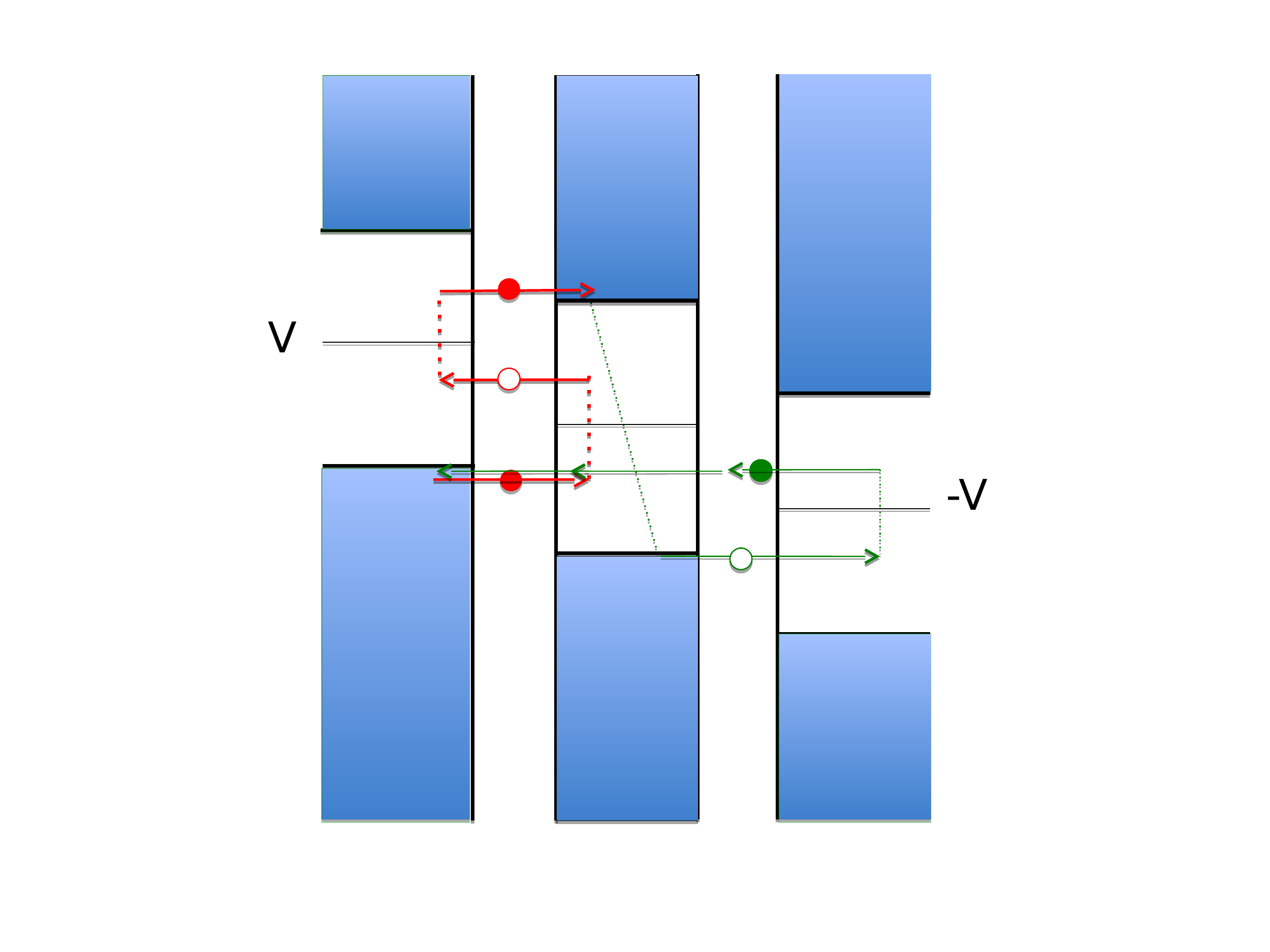}
\caption{(Color online.) Diagram of a typical phase-MAR process in the TTJ. 
The thick red arrow shows a quasiparticle path for the $3^\text{rd}$-order MAR between $S_1$ and $S_2$. 
The thin green arrow shows a multi-particle path for a 
multi-pair process that involves all superconductors by a cross-Andreev reflection in $S_2$. 
The resulting interference gives rise to phase-dependent MAR.}
\label{MAR_Quartet_BJ}
\end{center}
\end{figure}

\subsection{NDOS: phase dependence}

Unique to multi-terminal junctions is the strong dependence of the NDOS on the 
stationary phases of the superconductors. In particular, for the TTJ under 
the commensurate bias chosen, the NDOS depends
on the quartet phase, $\varphi_Q$, introduced in Sec. \ref{phenomenology}. Here we discuss
at length the effect of $\varphi_Q$, that we consider the central finding of this study.

 

\begin{figure}[t]
 \includegraphics[width=15pc]{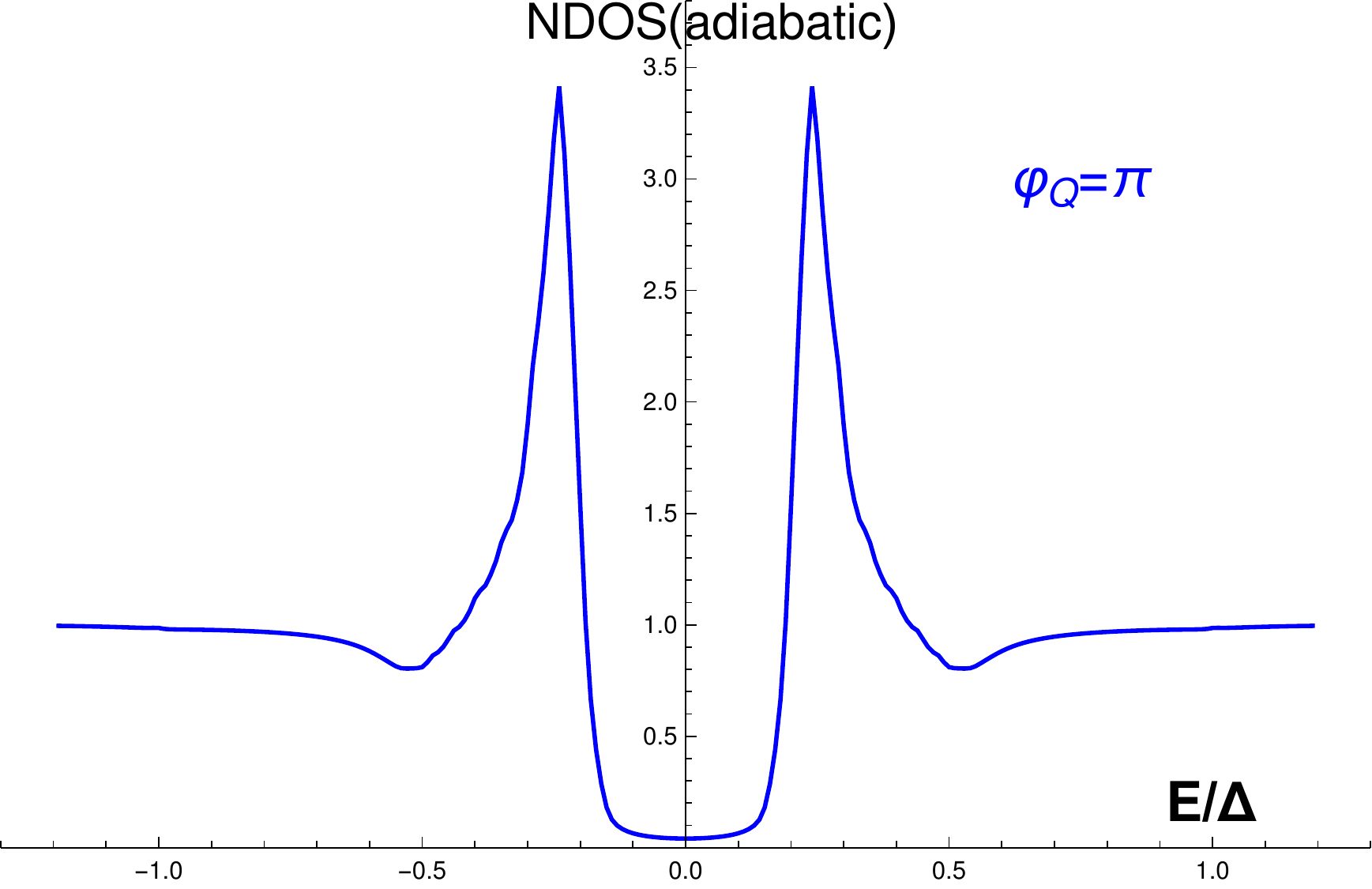}
 \caption{(Color online.) NDOS computed using an adiabatic approximation for small voltage, for a quartet phase $\varphi_Q=\pi$
 (see text for detail)}
 \label{fig:NDOS_adiabatic}
\end{figure}

\begin{figure*}[t!]
 \centerline{\includegraphics[height=8pc]{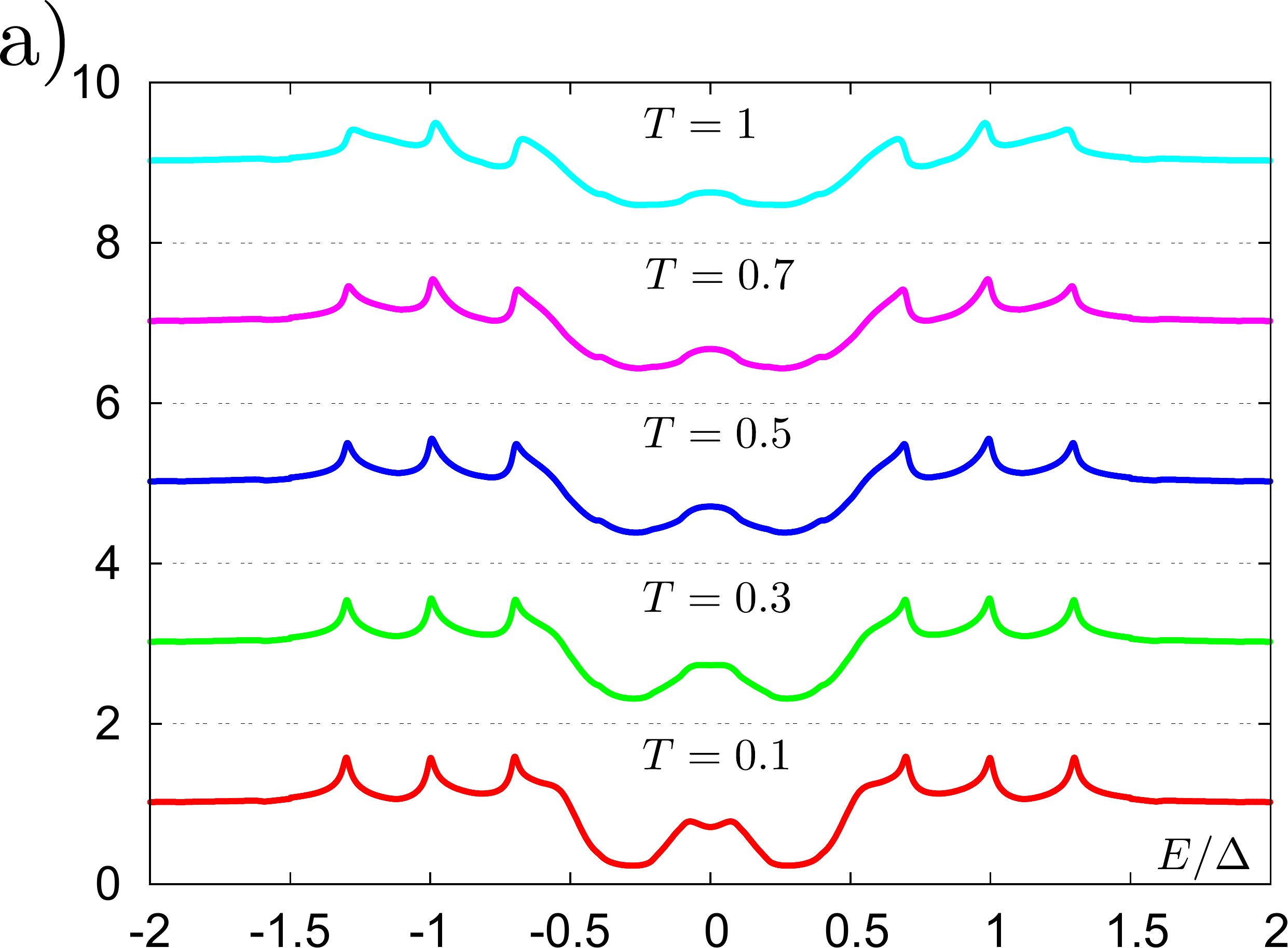} \hspace{1.0cm}
 \includegraphics[height=8pc]{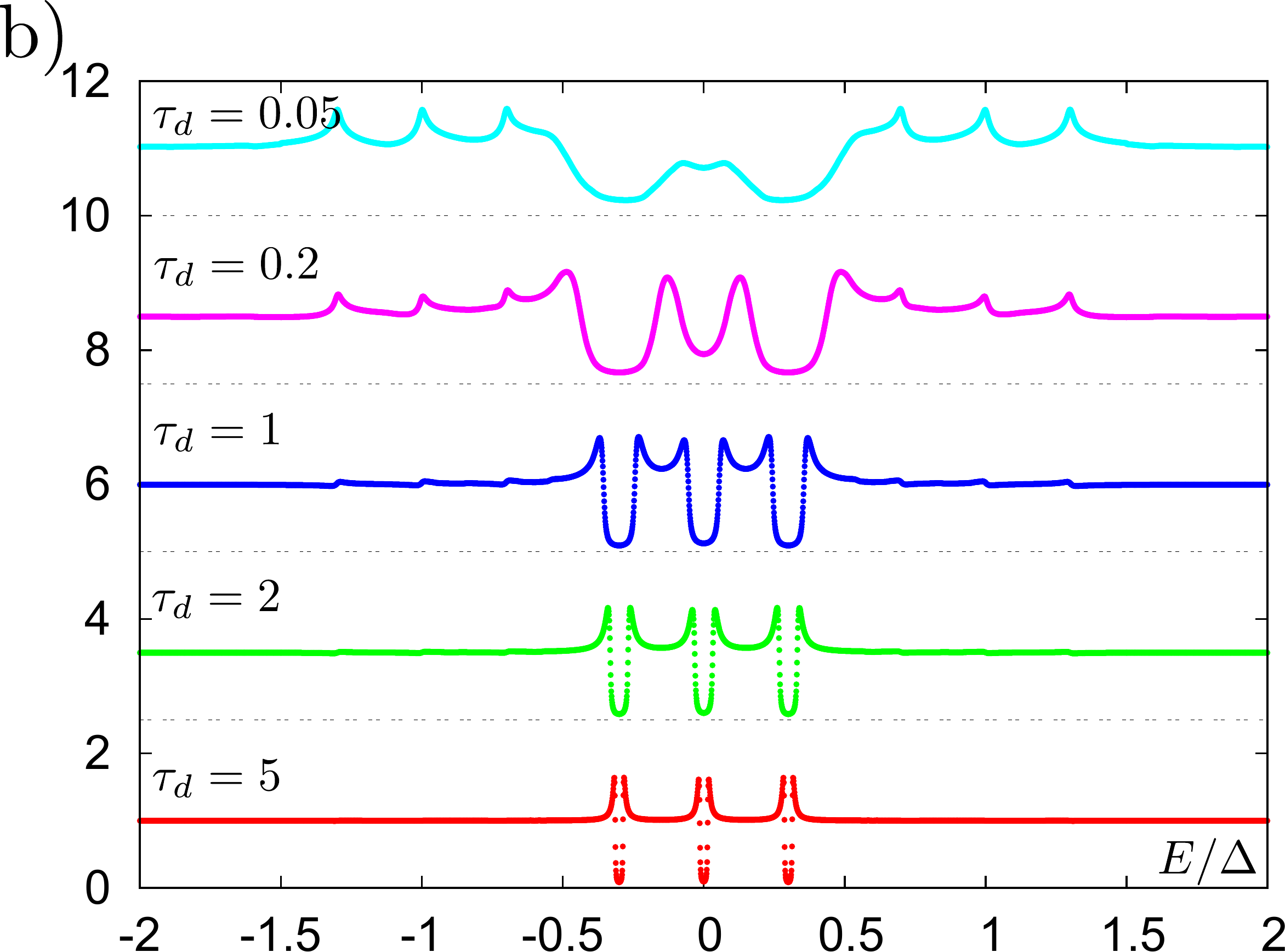} \hspace{1.0cm}
 \includegraphics[height=8pc]{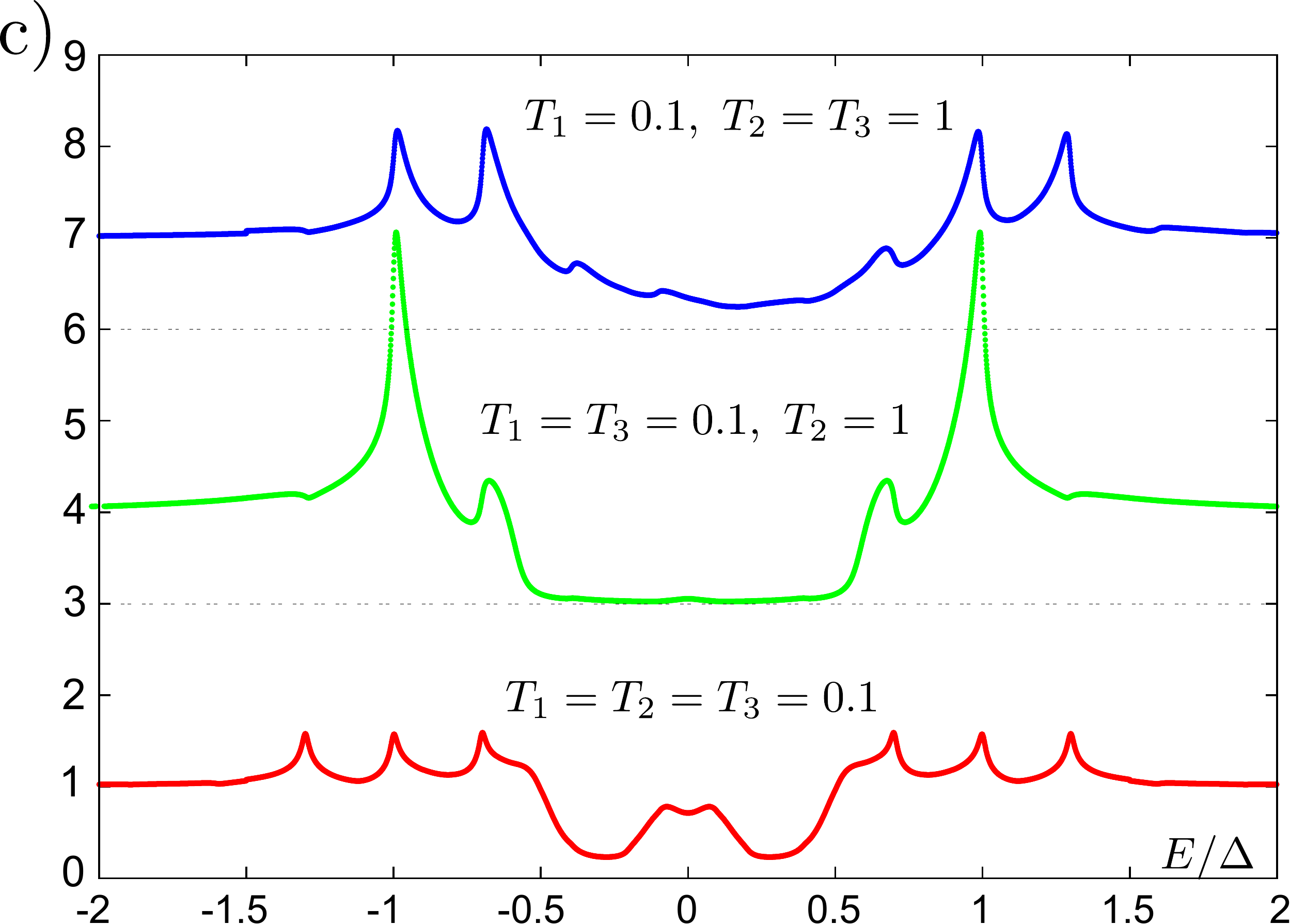}}
 \caption{(Color online.) Discussion of the pseudogap features of the NDOS
 for voltage $V=0.3$ and $\varphi_Q=\pi$, 
(a) for different transparencies in the symmetric junction, from the top $T=1, 0.7, 0.5, 0.3, 0.1$, 
(b) for different strengths of decoherence, from the top $\tau_d=0.05, 0.2, 1, 2, 5$, and
(c) for different transparency symmetries: top $T_1=0.1$, $T_2=T_3=1$; middle $T_1=T_3=0.1$, $T_2=1$; and bottom $T_1=T_2=T_3=0.1$.
The different curves are offset for clarity and the dotted horizontal lines indicate the offset, ${\cal N}(E)=0$.}
 \label{fig:NDOS_T_Ed}
\end{figure*}

As predicted in Sec. \ref{phenomenology}, the NDOS is a periodic function of $\varphi_Q$, with period $2 \pi$,
as can be seen in Figs.~\ref{fig:NDOS_PhiQ=Pi}(a) and \ref{fig:NDOS_PhiQ=Pi}(b). 
Pronounced differences between the NDOS at $\varphi_Q=0$ and $\varphi_Q=\pi$ can be seen in Fig.~\ref{fig:DOSbiased3SPhiPi}.
The height of the large peaks at $E = \pm \Delta, \pm (\Delta+V), \pm(\Delta-V)$ depends on $\varphi_Q$. 
While the position of these MAR resonances remains unchanged, the structure of the NDOS in-between the peaks is dramatically affected.
In particular, we focus on two dips in the NDOS, centered at $E=\pm V$, shown in Figs.~\ref{fig:DOSbiased3SPhiPi}(b) and \ref{fig:DOSbiased3SPhiPi}(c).
As a function of the quartet phase, the dips are shown in Fig.~\ref{fig:NDOS_PhiQ=Pi}(a) to be fully formed at $\varphi_Q=\pi$,
while disappearing at $\varphi_Q=0$.
Fixing the quartet phase at $\varphi_Q=\pi$, we show the dependence of the NDOS on voltage in Fig.~\ref{fig:NDOS_PhiQ=Pi}(c).
Upon lowering the voltage below the threshold $V=\Delta/2$, the dips emerge at $E= \pm V$. Upon further lowering the voltage,
the dips merge into a large, well defined pseudogap with sharp edges, seen in Fig.~\ref{fig:DOSbiased3SPhiPi}(d). 
Within the pseudogap the NDOS is not flat. 
An interference pattern develops characterized by small, isolated peaks in the NDOS that depend strongly on the voltage.
These small structures are diminished by decoherence and may be ascribed to local as well as non-local MAR processes.

The shape of the structures at $E=\pm V$ for $V\lesssim \Delta/2$ is similar to that of other dips
of the NDOS and does not remind of the sharp-edged shape of the bulk gap. 
However, they are unique in that they emerge upon the sharp threshold $V \leq \Delta/2$. 
The same threshold marks the limit where the quartet process can give rise to bound states in the junction, 
that do not couple to the continuum.
Such bound states support a non-dissipative transport of quasiparticle quartets. 
The structures at $E=\pm V$ are absent for $V=0.7$, see Fig.~\ref{fig:NDOS_PhiQ=Pi}(b).
Despite their shape, we consider it justified to refer to these dips at $E=\pm V$ 
as pseudogaps as they signal the presence of non-dissipative coherent processes. 


The variation with the phase $\varphi_Q$ of the amplitude and widths of the peaks in the NDOS
can be attributed to phase-dependent MAR processes and to the quartet process.  
The phase-dependent MAR processes~\cite{Jonckheere} can be understood by an interference 
between two distinct paths that transport a quasiparticle across the energy gap. 
An example is given in Fig.~\ref{MAR_Quartet_BJ}, where the third-order MAR process between 
$S_1$ and $S_2$ interferes with the process involving an Andreev reflection at the $S_2$-$S_3$ interface 
and a cross-Andreev reflection in $S_2$. The interference gives rise to the phase factor $e^{i(\varphi_1+\varphi_3)}$. 
The phase-MAR processes can be interpreted as binding, not only Cooper pairs to quasiparticles (as in usual MAR),
but also quartets to quasiparticles. 
	
The NDOS for $\varphi_Q=\pi$ at low voltage can be fitted by
the NDOS of a single superconducting electrode with a gap $\delta$, taken as the fitting parameter:
${\cal N}(E)=|x|/\sqrt{x^2-\delta^2}$ for $|x|>\delta$. The fit for $\delta=0.5$ is shown as a thin black dotted
line in Fig.~\ref{fig:DOSbiased3SPhiPi}(d). The fit reproduces the shape of the NDOS, 
except at the peaks, $E= \pm \Delta, \pm(\Delta + V), \pm(\Delta -V)$,
that are due to MAR processes. The curves for $V=0.1$ show that the NDOS resembles the NDOS of a junction
at equilibrium when $\varphi_Q=\pi$. 

A qualitative understanding can be obtained by appealing to the equilibrium NDOS, 
as calculated by the same model in Ref. \onlinecite{Padurariu} as a function of the two phases $\varphi_1, \varphi_3$, 
and measured in Ref. \onlinecite{Giazotto}. It is found that at low transparency, 
for a symmetric TTJ, the Andreev spectrum crosses zero energy in two points situated on the line $\varphi_Q=0$. 
Those regions extend in the phase plane for finite transparency, and the feature is robust against weak asymmetry. 
The proximity-induced minigap closes in some regions of the phase plane, 
particularly along the curve $\varphi_Q=0$, but remains open for $\varphi_Q=\pi$. 
At small voltage, in the adiabatic regime, we expect a minigap at $\varphi_Q=\pi$ and not at $\varphi_Q=0$. 
Qualitatively, the equilibrium Green’s function of the node is given by
$\check{G}_c(\varphi_1,\varphi_3)=\big(\omega-\check{\Sigma}(\omega)\big)^{-1}$. The self-energy can then be averaged 
to account for the rotation of the phase, $\varphi_1-\varphi_3=\frac{4eV}{\hbar}t$. 
This effective node Green's function $\langle\check{G}_c\rangle(\varphi_Q)$ depends only on the quartet phase. 
The corresponding NDOS is plotted in Fig.~\ref{fig:NDOS_adiabatic}, and matches qualitatively the exact calculation 
shown in Fig.~\ref{fig:DOSbiased3SPhiPi}(d). 
The sharp edges of the minigap are reproduced. The approximation is valid when the Andreev bound states change 
adiabatically with the variation of the phase, while preserving their occupation. For this reason, it cannot recover 
the exact peak amplitudes and the oscillations that reflect non-adiabatic features, 
e.g. Landau-Zener transitions between Andreev bound states and MAR processes.

	The effect of transmission, decoherence and asymmetry on the pseudogaps is
	explored in Fig. \ref{fig:NDOS_T_Ed}. Fig. \ref{fig:NDOS_T_Ed}(a) shows the relative
	robustness of the pseudogap structures as a function of transparency in the symmetric junction. 
	The position of the pseudogaps remains unchanged and only the depth reduces slightly
	as the transmission increases. 
	
	The dependence of the pseudogaps on the strength of decoherence is illuminating.
	Fig. \ref{fig:NDOS_T_Ed}(b) shows that at large decoherence the NDOS
	presents three minigaps, corresponding to three decoupled NS interfaces. As the
	decoherence strength decreases, the three minigaps broaden, and eventually,
	at $\tau_d \ll 1$, the three structures merge forming the two broad 
	pseudogaps observed for the coherent junction at $V=0.3$.

The origin of the pseudogap structures in the NDOS can be further probed by introducing an asymmetry
in the transmission of the contacts. Fig. \ref{fig:NDOS_T_Ed}(c) shows the situation when one of the contacts, 
or two of them, are fully transparent, while the other contacts have low transparency $T=0.1$. 
In this case, the two pseudogaps at $V=0.3$ are washed away. 
This is an additional argument for the importance of
non-local coherent transport processes in determining the structure of the NDOS. The dependence on 
symmetry suggests that pseudogaps originate from transport processes that involve all three superconductors.

\section{Conclusions.}
\label{conclusions}

We have presented a detailed study of the NDOS in a three-terminal Josephson junction under commensurate bias, $(V_1,V_2,V_3)=(V,0,-V)$. 

Using a phenomenological argument, we have shown that in general, the transport in this regime depends periodically on the phase combination, $\varphi_Q=(\varphi_1-\varphi_2)+(\varphi_3-\varphi_2)$, that is a constant of motion. The phase governs non-local transport of quartets, where the elementary transport process exchanges four quasiparticles between all three superconductors.
We argue that the quartet phase, $\varphi_Q$, can be tuned by the current flow in $S_2$.

Using the circuit theory formulation of non-equilibrium equations of superconducting transport, we reveal the NDOS, first for a conventional two-terminal junction, for comparison and as a benchmark of the method, and subsequently for the three-terminal junction, discussing its features for a wide range of parameters.

The complicated structure of the NDOS is greatly affected by electron-hole decoherence.
For small decoherence, the two- and three-terminal junctions have in common a complicated structure of peaks that we have shown to correspond to MAR resonances. We resolve the position of resonances in terms of bias and the order of MAR processes. The additional terminal gives rise to additional MAR channels in the three-terminal junction.

For large decoherence, we have shown that each NS interface contributes a minigap structure positioned around the chemical potential of the terminal. The interfaces appear decoupled in first approximation, with corrections arising due to finite decoherence. We have identified two transport regimes separated by the bias threshold $V=\Delta/2$. For low bias, the weaker coupling to the quasiparticle continuum gives rise to sharp features, while for high bias the coupling to the continuum is strong and the sharp edges are rounded.

The central result of our study is the dependence of the NDOS in the three-terminal junction on $\varphi_Q$. At $\varphi_Q=\pi$ and at small bias voltage the NDOS presents a sharp-edged pseudogap that forms around $E=0$. The sharp pseudogap bifurcates into a pair of pseudogaps for increasing bias, before vanishing for $V>\Delta/2$. Our predictions for the NDOS can be verified experimentally by tunneling spectroscopy. The presence of pseudogaps and the possibility to tune the NDOS by controlling the stationary phase in the presence of voltage bias has no equivalent in the two-terminal device, as it originates from non-local transport processes involving all three superconductors. In a subsequent contribution we will supplement this analysis of the NDOS by a discussion of the current flowing in the junction, focusing on the non-local and non-dissipative current that is discussed as a function of phase $\varphi_Q$ and voltage $V$.
The tunable and non-local properties of the coherent transport supported by the three-terminal junction recommend it as an interesting superconducting circuit element that may inspire quantum engineering applications.

\begin{acknowledgments}

We gratefully acknowledge the fruitful exchange of ideas with our colleague R\'egis M\'elin.

We acknowledge support from the French National Research Agency (ANR) through the project ANR-Nano-Quartets (ANR-12-BS1000701). 
This work has been carried out in the framework of the Labex Archim\'ede (ANR-11-LABX-0033) 
and of the A*MIDEX project (ANR-11-IDEX-0001-02), funded by the \enquote{Investissements d'Avenir} 
French Government program managed by ANR. C.P. acknowledges support from the Academy of Finland and from the Centre for Quantum Engineering at Aalto University.

\end{acknowledgments}

\appendix*
\section{Fourier transforms}

Green's functions generally depend on two time-space coordinates $\check{G}(\mathbf{x_1},\mathbf{x_2},t_1,t_2)$. In the circuit theory model the junction is separated into regions where transport is uniform and can be characterized by a coordinate independent Green's function, $\check{G}(t_1,t_2)$. These are the Green's functions that describe the terminals and the Green's function of the central node.

The Green's functions are related by equations originating from the Usadel equation where matrix products are defined in the Keldysh-Nambu-$t_1$-$t_2$ space. Since the time indices are continuous, matrix products take the form of a convolution in time, $\check{G} = \check{G}_1 \circ \check{G}_2$,
\begin{align}
\check{G}(t_1,t_2) = \int dt\ \check{G}_1(t_1,t)\check{G}_2(t,t_2).
\end{align}
In energy representation,
\begin{align}
\check{G}(E_1,E_2) &= \int dt_1dt_2\ e^{i(E_1t_1-E_2t_2)}\check{G}(t_1,t_2),\\
\check{G}(E_1,E_2) &= \int \frac{dE}{2\pi}\ \check{G}_1(E_1,E)\check{G}_2(E,E_2).
\end{align}
Under stationary transport conditions the Green's function depends only on $(t_1-t_2)$, or alternatively, only on one energy $G(E_1,E_2)=G(E_1)\delta(E_1-E_2)$. In this case, the convolution reduces to a simple matrix product of Keldysh-Nambu Green's functions defined at a given energy. The equations are therefore easily implemented numerically.

Our work addresses transport in non-stationary conditions. As discussed in Sec.~\ref{microscopic}, the transport is periodic and described by the Josephson frequency $\omega_0=2eV/\hbar$. We define the following double time Fourier transform to a representation in energy and harmonics of $\omega_0$, (here we have set $\hbar=1$)
\begin{widetext}
\begin{align}
\check{G}(E,n,m) &= \frac{1}{T}\displaystyle\int^{\infty}_{-\infty} dt_1 \int^{T-t_1}_{-T-t_1} dt_2\ e^{i(E+n\omega_0)t_1}e^{-i(E+m\omega_0)t_2}\ \check{G}(t_1,t_2), \quad T=\frac{2\pi}{\omega_0}
\end{align}
\end{widetext}
The restriction in integration over $t_2$ implements the condition, $-T/2<(t_1+t_2)/2<T/2$, valid due to periodicity, $\check{G}(t_1,t_2)=\check{G}(t_1+T,t_2+T)$.

With the above transformation it can be shown that the convolution product becomes a product of matrices defined at a given energy $E$, with size corresponding to the Keldysh-Nambu-$n$-$m$ space, $\check{G} = \check{G}_1 \circ \check{G}_2$,
\begin{align}
\check{G}(E,n,m) = \displaystyle \sum_p \check{G}_1(E,n,p) \check{G}_2(E,p,m).
\end{align}
We have used this expression to find the numerical solution of the non-stationary transport problem, further detailed in Sec.~\ref{microscopic}.

\end{document}